\documentclass[twocolumn,amsmath,amssymb,prd]{revtex4}

\def\al{\alpha}
\def\be{\beta}
\def\ga{\gamma}
\def\de{\delta}
\def\ep{\epsilon}

\def\et{\eta}

\def\ka{\kappa}
\def\la{\lambda}

\def\rh{\rho}

\def\si{\sigma}

\def\ta{\tau}

\def\ch{\chi}

\def\om{\omega}

\def\La{\Lambda}

\def\Ph{\Phi}

\def\cE{{\cal E}}
\def\cl{{\cal L}}
\def\cL{{\cal L}}

\def\mn{{\mu\nu}}

\def\fr#1#2{{{#1} \over {#2}}}
\def\half{{\textstyle{1\over 2}}}

\def\frac#1#2{{\textstyle{{#1}\over {#2}}}}

\def\vev#1{\langle {#1}\rangle}

\def\lsim{\mathrel{\rlap{\lower4pt\hbox{\hskip1pt$\sim$}}
     \raise1pt\hbox{$<$}}}
\def\gsim{\mathrel{\rlap{\lower4pt\hbox{\hskip1pt$\sim$}}
     \raise1pt\hbox{$>$}}}
\def\sqr#1#2{{\vcenter{\vbox{\hrule height.#2pt
          \hbox{\vrule width.#2pt height#1pt \kern#1pt
          \vrule width.#2pt}
          \hrule height.#2pt}}}}
\def\square{\mathchoice\sqr66\sqr66\sqr{2.1}3\sqr{1.5}3 \,}

\def\prt{\partial}

\def\etal{{\it et al.}}

\def\pt#1{\phantom{#1}}

\def\ol#1{\overline{#1}}

\def\nsc#1#2#3{\om_{#1}^{{\pt{#1}}#2#3}}

\def\vb#1#2{e_{#1}^{{\pt{#1}}#2}}
\def\ivb#1#2{e^{#1}_{{\pt{#1}}#2}}
\def\uvb#1#2{e^{#1#2}}
\def\lvb#1#2{e_{#1#2}}

\def\son{\si_1}
\def\stw{\si_2}
\def\ton{\ta_1}
\def\ttw{\ta_2}
\def\tth{\ta_3}
\def\ks{KS}

\newcommand{\beq}{\begin{equation}}
\newcommand{\eeq}{\end{equation}}
\newcommand{\bea}{\begin{eqnarray}}
\newcommand{\eea}{\end{eqnarray}}
\newcommand{\bit}{\begin{itemize}}
\newcommand{\eit}{\end{itemize}}
\newcommand{\rf}[1]{(\ref{#1})}

\begin{document}

\title{
Spontaneous Lorentz and Diffeomorphism Violation, 
Massive Modes, and Gravity}

\author{Robert Bluhm$^a$, 
Shu-Hong Fung$^{a,b}$,
and V.\ Alan Kosteleck\'y$^c$}

\affiliation{
$^a$Physics Department, Colby College,
Waterville, ME 04901
\\
$^b$Physics Department, Chinese University of Hong Kong,
Shatin, N.T., Hong Kong
$^c$Physics Department, Indiana University,
Bloomington, IN 47405
}

\date{IUHET 509, December 2007;
accepted in Physical Review D}

\begin{abstract}
Theories with spontaneous local Lorentz and diffeomorphism violation  
contain massless Nambu-Goldstone modes,
which arise as field excitations in the minimum 
of the symmetry-breaking potential.
If the shape of the potential also allows excitations 
above the minimum,
then an alternative gravitational Higgs mechanism can occur
in which massive modes involving the metric appear.
The origin and basic properties of the massive modes
are addressed in the general context 
involving an arbitrary tensor vacuum value.
Special attention is given to the case of bumblebee models,
which are gravitationally coupled vector theories 
with spontaneous local Lorentz and diffeomorphism violation.
Mode expansions are presented in both local and spacetime frames,
revealing the Nambu-Goldstone and massive modes
via decomposition of the metric and bumblebee fields,
and the associated symmetry properties and gauge fixing 
are discussed.
The class of bumblebee models with kinetic terms
of the Maxwell form is used as a focus for more detailed study.
The nature of the associated conservation laws 
and the interpretation as a candidate alternative 
to Einstein-Maxwell theory are investigated. 
Explicit examples 
involving smooth and Lagrange-multiplier potentials
are studied to illustrate features of the massive modes,
including their origin, nature, dispersion laws, 
and effects on gravitational interactions. 
In the weak static limit,
the massive mode and Lagrange-multiplier fields are found 
to modify the Newton and Coulomb potentials.
The nature and implications of these modifications are examined.
\end{abstract}


\maketitle

\section{Introduction}
\label{Introduction}

In relativistic quantum field theory,
the nature of the field modes associated
with the spontaneous breaking of an internal symmetry
is now standard lore.
When a global internal symmetry is spontaneously broken,
one or more massless modes called Nambu-Goldstone (NG) modes appear 
\cite{ng}.
If instead the symmetry is local,
the Higgs mechanism can occur:
the massless NG modes play the role 
of additional components of the gauge fields,
which then propagate as massive modes
\cite{hm}.
In either case,
the spontaneous symmetry breaking
is typically driven by a potential term $V$ 
in the Lagrange density.
The vacuum field configuration lies 
in a minimum $V_0$ of $V$.
The massless NG modes can be understood
as field excitations about the vacuum 
that preserve the value $V_0$,
and they are associated with the broken generators of the symmetry.
For many potentials,
there are also additional excitations 
involving other values of $V$.
These excitations,
often called Higgs modes,
correspond to additional massive modes
that are distinct from any massive gauge fields.

This standard picture changes when the spontaneous breaking
involves a spacetime symmetry rather than an internal one.
In this work, 
we focus on spontaneous breaking 
of Lorentz and diffeomorphism symmetries,
for which the corresponding Higgs mechanisms
exhibit some unique features
\cite{ks}.
Spontaneous Lorentz violation occurs 
when one or more Lorentz-nonsinglet field configurations
acquire nonzero vacuum expectation values.
The field configurations of interest
can include fundamental vectors or tensors,
derivatives of scalars and other fields,
and Lorentz-nonsinglet composites.
The nonzero vacuum values are manifest 
both on the spacetime manifold and in local frames 
\cite{akgrav}.
Their origin in spontaneous violation 
implies both local Lorentz violation and diffeomorphism violation,
along with the existence of NG modes
\cite{bk}.

At the level of an underlying Planck-scale theory,
numerous proposals exist that involve
spontaneous Lorentz violation, 
including ones based on
string theory
\cite{ksp},
noncommutative field theories
\cite{ncqed},
spacetime-varying fields
\cite{spacetimevarying},
quantum gravity
\cite{qg},
random-dynamics models
\cite{fn},
multiverses
\cite{bj},
brane-world scenarios
\cite{brane},
supersymmetry
\cite{susy},
and massive gravity 
\cite{modgrav1,modgrav2,modgrav3}.
At experimentally accessible scales,
the observable signals resulting from Lorentz breaking
can be described using effective field theory
\cite{kp}.
The general realistic effective field theory
containing the Lagrange densities 
for both general relativity and the Standard Model
along with all scalar terms 
involving operators for Lorentz violation
is called the Standard-Model Extension (SME)
\cite{ck,akgrav}.
Searches for low-energy signals of Lorentz violation 
represent a promising avenue of investigation involving 
the phenomenology of quantum gravity
\cite{cpt07}.
Numerous experimental measurements of SME coefficients
for Lorentz violation have already been performed \cite{kr}, 
including ones with
photons \cite{photonexpt},
electrons \cite{eexpt,eexpt2,eexpt3},
protons and neutrons \cite{ccexpt,spaceexpt,bnsyn},
mesons \cite{hadronexpt},
muons \cite{muexpt},
neutrinos \cite{nuexpt},
the Higgs \cite{higgs},
and gravity \cite{gravexpt,bak06}. 

For spontaneous Lorentz and diffeomorphism breaking,
a general analysis 
of the nature of the NG modes and Higgs mechanisms 
is provided in Ref.\ \cite{bk}.
One result 
is that the spontaneous breaking
of local Lorentz symmetry 
implies spontaneous breaking of diffeomorphism symmetry
and vice versa.
Since six local Lorentz transformations and four diffeomorphisms
can be broken,
up to ten NG modes can appear.
To characterize these,
it is natural to adopt the vierbein formalism
\cite{uk},
in which the roles of local Lorentz transformations
and diffeomorphisms are cleanly distinguished.
It turns out that the vierbein itself
naturally incorporates all ten modes.
In an appropriate gauge,
the six Lorentz NG modes appear 
in the antisymmetric components of the vierbein,
while the four diffeomorphism NG modes 
appear along with the usual gravitational modes
in the symmetric components.

The dynamical behavior of the various NG modes is determined
by the structure of the action 
\cite{bk}.
In a Lagrange density formed from tensor quantities
and with diffeomorphism-covariant kinetic terms,
the diffeomorphism NG modes are nonpropagating.
This feature, unique to diffeomorphism symmetry,
can be viewed as arising 
because the diffeomorphism NG field excitations 
that preserve $V_0$ include metric excitations,
and the combined excitations cancel 
at propagation order in covariant derivatives and curvature.
In contrast,
the number of propagating Lorentz NG modes
is strongly model dependent.
For example,
choosing the kinetic terms in the Lagrange density
of the original field theory
to eliminate possible ghost modes
can also prevent the propagation of one or more Lorentz NG modes,
leaving them instead as auxiliary fields.

Several types of possible Higgs mechanisms can be distinguished 
when spacetime symmetries are spontaneously broken.
These have features distinct from the 
conventional Higgs mechanism of gauge field theories
\cite{hm}
and Higgs mechanisms involving gravity without
Lorentz violation
\cite{hmnonlv}.
The analysis of Higgs mechanisms can be performed 
either using the vierbein formalism,
which permits tracking of Lorentz and diffeomorphism properties,
or by working directly with fields on the spacetime manifold.
The results of both approaches are equivalent.

For local Lorentz symmetry,
the role of the gauge fields in the vierbein formalism 
is played by the spin connection.
The \it Lorentz Higgs mechanism \rm
occurs when the Lorentz NG modes 
play the role of extra components of the spin connection
\cite{bk}.
Some components of the spin connection then acquire mass
via the covariant derivatives in the kinetic
part of the Lagrange density.
Explicit models displaying the Lorentz Higgs mechanism are known.
For this mechanism to occur,
the components of the spin connection 
must propagate as independent degrees of freedom,
which requires a theory based on Riemann-Cartan geometry.
If instead the theory is based on Riemann geometry,
like General Relativity,
the spin connection is fixed nonlinearly 
in terms of the vierbein and its derivatives.
The presence of these derivatives ensures 
that no mass terms emerge 
from the kinetic part of the Lagrange density,
although the vierbein propagator is modified.

In the context of diffeomorphism symmetry,
the role of the gauge fields is played by the metric.
For spontaneous diffeomorphism breaking in Riemann spacetime,
a conventional 
\it diffeomorphism Higgs mechanism \rm
cannot generate a mass for the graviton
because the connection 
and hence the analogue of the usual mass term
involve derivatives of the metric
\cite{ks}.
Also,
since diffeomorphism NG modes are nonpropagating 
in a Lagrange density with covariant kinetic terms
for reasons mentioned above,
the propagating NG degrees of freedom 
required to generate massive fields via a
conventional diffeomorphism Higgs mechanism are lacking. 

In a conventional gauge theory with a nonderivative potential $V$,
the gauge fields are absent from $V$
and so the potential cannot directly contribute to the gauge masses. 
However,
in spontaneous Lorentz and diffeomorphism violation,
massive Higgs-type modes involving the vierbein can 
arise via the \it alternative Higgs mechanism, \rm 
which involves the potential $V$ 
\cite{ks}.
The key point is that $V$ contains both tensor and metric fields,
so field fluctuations about the vacuum value $V_0$
can contain quadratic mass terms involving the metric.

In this work,
we study the nature and properties 
of the additional massive Higgs-type modes
arising from this alternative Higgs mechanism.
A general treatment is provided
for a variety of types of potentials $V$
in gravitationally coupled theories with Riemann geometry.
We investigate the effects of the massive modes 
on the physical properties of gravity.
In certain theories with spontaneous Lorentz violation,
the NG modes can play the role of photons
\cite{bk},
and we also examine the effects of the massive modes 
on electrodynamics in this context.

The next section of this work provides a general discussion
of the origin and basic properties of the massive Higgs-type modes.
Section 
\ref{Bumblebee Models}
analyzes the role of these modes 
in vector theories with spontaneous Lorentz violation,
known as bumblebee models.
In Sec.\ \ref{Examples},
the massive modes in
a special class of bumblebee models are studied in more detail
for several choices of potential in the Lagrange density,
and their effects
on both the gravitational and electromagnetic interactions
are explored.
Section \ref{Summary} summarizes our results.
Some details about transformation laws are provided in the 
Appendix. 
Throughout this work,
the conventions and notations of Refs.\ \cite{akgrav,bk} 
are used.

\section{Massive Modes}
\label{Massive Modes}

The characteristics of the massive Higgs-type modes 
that can arise from the alternative Higgs mechanism
depend on several factors.
Among them are 
the type of field configuration
acquiring the vacuum expectation value
and the form of the Lagrange density,
including the choice of potential $V$ 
inducing spontaneous breaking 
of Lorentz and diffeomorphism symmetries.
In this section,
we outline some generic features associated with 
the alternative Higgs mechanism
in a theory of a general tensor field $T_{\la\mu\nu\cdots}$
in a Riemann spacetime with metric $g_{\mu\nu}$.
We consider first consequences 
of the choice of potential $V$,
then discuss properties of vacuum excitations,
and finally offer comments on the massive modes
arising from the alternative Higgs mechanism.

\subsection{Potentials}
\label{Potentials}

The potential $V$ in the Lagrange density is taken 
to trigger a nonzero vacuum expectation value 
\beq
\vev{T_{\la\mu\nu\cdots}}=
t_{\la\mu\nu\cdots} 
\label{Tvev}
\eeq
for the tensor field,
thereby spontaneously breaking
local Lorentz and diffeomorphism symmetries.
In general, 
$V$ varies with $T_{\la\mu\nu\cdots}$, 
its covariant derivatives,
the metric $g_{\mu\nu}$,
and possibly other fields.
However,
for simplicity we suppose here that
$V$ has no derivative couplings
and involves only $T_{\la\mu\nu\cdots}$
and $g_{\mu\nu}$.
We also suppose that $V$ is everywhere positive
except at its degenerate minima,
which have $V = 0$ and are continuously connected
via the broken Lorentz and diffeomorphism generators.
The vacuum is chosen to be the particular minimum
in which $T_{\la\mu\nu\cdots}$ attains
its nonzero value \rf{Tvev}.

Since the Lagrange density is an observer scalar,
$V$ must depend on fully contracted combinations 
of $T_{\la\mu\nu\cdots}$ and $g_{\mu\nu}$.
Provided $T_{\la\mu\nu\cdots}$ has finite rank,
the number of independent scalar combinations is limited. 
For example,
for a symmetric two-tensor field $C_{\mu\nu}$
there are four independent possibilities, 
which can be given explicitly 
in terms of traces of powers of $C_{\mu\nu}$ and $g_{\mu\nu}$ 
\cite{kpgr}.
It is convenient to denote generically 
these scalar combinations as $X_m$,
where $m= 1,2,\ldots, M$ ranges over the number $M$
of independent combinations.
The functional form of the potential therefore takes the form
\beq
V = V(X_1,X_2,\ldots,X_M)
\eeq
in terms of the scalars $X_m$.

The definition of the scalars $X_m$
can be chosen so that $X_m = 0$ in the vacuum for all $m$.
For example,
a choice involving a quadratic combination of the tensor 
with zero vacuum value is
\beq
X =
T_{\la\mu\nu\cdots}
g^{\la\al} g^{\mu\be} g^{\nu\ga} \cdots
T_{\al\be\ga\cdots} \pm t^2,
\label{X}
\eeq
with $t^2$ the norm 
\beq
t^2 = 
\mp t_{\la\mu\nu\cdots}
\vev{g^{\la\al}} \vev{g^{\mu\be}} \vev{g^{\nu\ga}} \cdots
t_{\al\be\ga\cdots} ,
\label{norm}
\eeq
where $\vev{g^{\la\al}}$ 
is the vacuum value of the inverse metric.
The $\mp$ sign is introduced for convenience 
so that $t^2$ can be chosen nonnegative.
In principle,
$t^2$ could vary with spacetime position,
which would introduce explicit spacetime-symmetry breaking,
but it suffices for present purposes
to take $t^2$ as a real nonnegative constant.

The $M$ conditions $X_m = 0$ fix the vacuum value
$t_{\la\mu\nu\cdots}$.
If only a subset of $N$ of these $M$ conditions
is generated in a given theory, 
then the value of $t_{\la\mu\nu\cdots}$ is specified
up to $(M-N)$ coset transformations,
and so the vacuum is degenerate under $(M-N)$
additional gauge symmetries. 
Note that these $(M-N)$ freedoms are distinct
from Lorentz and diffeomorphism transformations.

It is useful to distinguish two classes of potentials $V$.
The first consists of smooth functionals $V$ of $X_m$
that are minimized by the conditions $X_m=0$
for at least some $m$.
These potentials therefore satisfy $V = V_m^\prime = 0$
in the vacuum,
where $V_m^\prime$ denotes the derivative with respect to $X_m$,
and they fix the vacuum value of $T_{\la\mu\nu\cdots}$ 
to $t_{\la\mu\nu\cdots}$ modulo possible gauge freedoms.
A simple example with the quantity $X$ in Eq.\ \rf{X} is 
\beq
V_S(X) = \half \ka X^2,
\label{VS}
\eeq
where $\ka$ is a constant.
In this case,
the vacuum value $t_{\la\mu\nu\cdots}$ 
is a solution of $V = V^\prime = 0$,
where $V^\prime$ denotes a derivative with respect to $X$.
If the matrix $V_{mn}^{\prime\prime}$
of second derivatives has nonzero eigenvalues,
the smooth functionals $V$ can give rise
to quadratic mass terms in the Lagrange density 
involving the tensor and metric fields.
Potentials $V$ of this type are therefore associated 
with the alternative Higgs mechanism,
and they are the primary focus of our attention.

A second class of potentials 
introduces Lagrange-multiplier fields $\la_m$ 
for at least some $m$,
to impose directly the conditions $X_m=0$ 
as constraint terms in the Lagrange density.
We consider here both linear and quadratic functional
forms for these constraints.
An explicit linear example is 
\beq
V_L(\la, X) = \la X,
\label{VL}
\eeq
while a quadratic one is 
\beq 
V_Q(\la, X) = \half \la X^2.
\label{VQ}
\eeq
In each example,
$\la$ is a Lagrange-multiplier field
and has equation of motion with solution $X=0$,
so the value $t_{\la\mu\nu\cdots}$ 
is a vacuum solution for the tensor.
Potentials in the Lagrange-multiplier class are unlikely 
to be physical in detail 
because they enforce singular slicings 
in the phase space for the fields.
However,
when used with care they can be useful
as limiting approximations to potentials in the smooth class,
including those inducing the alternative Higgs mechanism
\cite{ks}.
Note that positivity of the potentials
can constrain the range of the Lagrange multiplier field.
For example,
the off-shell value of 
$\la$ in $V_L$ must have the same sign as $X$,
while that of $\la$ in $V_Q$ must be non-negative.

\subsection{Excitations}
\label{Excitations}

Field excitations about the vacuum solution \rf{Tvev}
can be classified in five types:
gauge modes, 
NG modes, 
massive modes,
Lagrange-multiplier modes,
or spectator modes.
Gauge modes arise if the potential $V$ fixes only  
$(M-N)$ of the $M$ conditions $X_m = 0$,
so that the vacuum is unspecified up to $N$ conditions.
These modes can be disregarded for most purposes here 
because they can be eliminated via gauge fixing
without affecting the physics.
The NG modes are generated by the virtual action
of the broken Lorentz and diffeomorphism generators
on the symmetry-breaking vacuum.
They preserve the vacuum condition $V=0$.
Massive modes are excitations for which the symmetry breaking
generates quadratic mass terms.
Lagrange-multiplier modes are excitations
of the Lagrange multiplier field.
Finally,
spectator modes are the remaining modes in the theory.

For smooth potentials,
field excitations preserving the conditions $X_m = 0$ for all $m$ 
have potential $V = 0$.
They also satisfy $V_m^\prime = 0$.
The NG modes are of this type.
Excitations with $X_m \neq 0$ 
having nonzero potential $V \neq 0$
and $V_m^\prime \neq 0$
are massive modes,
with mass matrix determined by 
the second derivatives $V_{mn}^{\prime\prime}$.
Since the smooth potentials depend
on the tensor $T_{\la\mu\nu\cdots}$
and the metric $g_{\mu\nu}$,
the corresponding massive modes 
involve combinations of excitations of these fields.

In contrast,
for Lagrange-multiplier potentials 
the conditions $X_m = 0$ always hold on shell,
which implies $V=0$ for all on-shell excitations.
If also $V_m^\prime = 0$, 
then the excitations remain in the potential minimum
and include the NG modes.
For linear functional forms of $V$,
it follows that $V_m^\prime = \la_m$,
so $V_m^\prime$ is nonzero
only when the Lagrange-multiplier field $\la_m$ is excited.
Any excitations of $T_{\la\mu\nu\cdots}$ and $g_{\mu\nu}$
must have $V = V_m^\prime = 0$.
For quadratic functional forms of $V$,
one finds $V_m^\prime = 0$ for all excitations,
including the $\la$ field.
We therefore can conclude
that the combinations of $T_{\la\mu\nu\cdots}$ and $g_{\mu\nu}$
playing the role of massive modes for smooth potentials
are constrained to zero for Lagrange-multiplier potentials.
Evidently,
studies of the alternative Higgs mechanism  
must be approached with care
when the Lagrange-multiplier approximation to a smooth potential
is adopted.
 
The tensor excitations about the vacuum can be expressed
by expanding $T_{\la\mu\nu\cdots}$ as
\beq
T_{\la\mu\nu\cdots} =
t_{\la\mu\nu\cdots} + \ta_{\la\mu\nu\cdots},
\label{tau}
\eeq
where the excitation 
$\ta_{\la\mu\nu\cdots}$
is defined as the difference 
$\ta_{\la\mu\nu\cdots} \equiv
\de T_{\la\mu\nu\cdots}
= T_{\la\mu\nu\cdots} - t_{\la\mu\nu\cdots}$
between the tensor and its vacuum value.
We also expand the metric 
\beq
g_{\mu\nu} = \vev{g_{\mu\nu}} + h_{\mu\nu} 
\label{ghmunu}
\eeq
in terms of the metric excitations $h_{\mu\nu}$ 
about the metric background value 
$\vev{g_{\mu\nu}}$.
For simplicity and definiteness,
in much of what follows we take 
the background metric to be that of Minkowski spacetime,
$\vev{g_{\mu\nu}} = \et_{\mu\nu}$.
We also suppose $t_{\la\mu\nu\cdots}$ is constant
in this background,
so that 
$\prt_\al t_{\la\mu\nu\cdots} = 0$
in cartesian coordinates.

Other choices can be made for the expansion of the tensor
about its vacuum value. 
One alternative 
is to expand the contravariant version of the tensor as
\beq
T^{\la\mu\nu\cdots} = 
\overline t^{\la\mu\nu\cdots} + \widetilde T^{\la\mu\nu\cdots}.
\label{olt}
\eeq
In a Minkowski background,
the vacuum values in the two expansions \rf{tau} and \rf{olt}
are related simply by
\beq
\overline t^{\la\mu\nu\cdots} 
= \et^{\la\al}\et^{\mu\be}\et^{\nu\ga}\cdots 
t_{\al\be\ga\cdots}.
\eeq
However,
the relationship between the two tensor excitations 
at leading order involves also the metric excitation:
\beq
\widetilde T^{\la\mu\nu\cdots}
=
\ta^{\la\mu\nu\cdots}
- h^{\la\al}t_\al^{\pt{\al}\mu\nu\cdots}
- h^{\mu\al}t^{\la\pt{\al}\nu\cdots}_{\pt{\la}\al}
- h^{\nu\al}t^{\la\mu\pt{\al}\cdots}_{\pt{\la\mu}\al}
- \ldots.
\eeq
In this expression,
indices have been raised using the Minkowski metric.

Any Lagrange multipliers $\la_m$ in the theory
can also be expanded about their vacuum values 
$\overline\la_m$
as 
\beq
\la_m = \overline\la_m + \widetilde\la_m .
\eeq
For linear Lagrange-multiplier potentials,
the equations of motion for 
$T_{\la\mu\nu\cdots}$ and $g_{\mu\nu}$
provide constraints on the vacuum values $\overline\la_m$.
In a Minkowski background 
and a potential yielding $X_m = 0$ for all $m$,
the equations of motion for $T_{\la\mu\nu\cdots}$ 
can be solved to give 
\beq
\overline\la_m = 0.
\eeq 
For quadratic Lagrange-multiplier potentials,
the $\la_m$ are absent from the equations of motion.
Their vacuum values are therefore physically irrelevant,
and $\overline\la_m = 0$ can also be adopted in this case.
For the remainder of this work
we take $\overline\la_m = 0$,
and for notational simplicity
we write $\la_m$ for both 
the full field $\la_m$ and the excitation $\widetilde\la_m$.

\subsection{Massive modes}
\label{Massive modes}

For a smooth potential $V$,
the massive excitations typically 
involve a mixture of tensor and gravitational fields.
As an example,
consider the expression for $V_S$ in Eq.\ \rf{VS}.
This can be expanded in terms of the excitations 
$\ta_{\la\mu\nu\cdots}$ and $h_{\mu\nu}$.
Retaining only terms up to quadratic order 
in a Minkowski background gives 
\bea
V_S &\approx& 2 \ka [ t^{\la\mu\nu\cdots} (\ta_{\la\mu\nu\cdots}
- \half h_{\la\al} t^\al_{\pt{\al}\mu\nu\cdots}
- \half h_{\mu\be} t_{\la\pt{\be}\nu\cdots}^{\pt{\la}\be}
\nonumber\\
&&
\qquad
\qquad
\qquad
\quad
- \half h_{\nu\ga} t_{\la\mu\pt{\ga}\cdots}^{\pt{\la\mu}\ga}
- \ldots)]^2 ,
\label{Vexpans}
\eea
where index contractions are performed
with the Minkowski metric $\et_{\mu\nu}$.
Evidently,
the massive excitations in this example
involve linear combinations of $\ta_{\la\mu\nu\cdots}$ 
with contractions of $h_{\mu\nu}$ and $t_{\la\mu\nu\cdots}$.

The explicit expressions for the massive modes
can be modified by local Lorentz and diffeomorphism gauge fixing.
The action is symmetric 
under 10 local Lorentz and diffeomorphism symmetries,
so there are 10 possible gauge conditions.
For an unbroken symmetry generator,
the effects of a gauge choice are conventional.
For a broken symmetry generator,
a gauge choice fixes the location in field space
of the corresponding NG mode.
For example,
suitable gauge choices can place all the NG modes
in the vierbein
\cite{bk}.
These gauge choices also affect the form of the massive modes. 
They can be used to isolate some or all of the massive modes 
as components of either the gravitational or tensor fields
by gauging other components to zero.
Alternatively,
the gauge freedom can be used to
simplify the equations of motion,
while the massive modes remain a mixture
of the excitations $\ta_{\la\mu\nu\cdots}$ and $h_{\mu\nu}$.

The behavior of the massive modes depends
on the form of the kinetic terms in the Lagrange density 
as well as the form of the potential $V$.
A gravitational theory with a dynamical tensor field
$T_{\la\mu\nu\cdots}$
has kinetic terms for both
$g_{\mu\nu}$ and $T_{\la\mu\nu\cdots}$ 
and hence for both 
$h_{\mu\nu}$ and $\ta_{\la\mu\nu\cdots}$.
Since in the alternative Higgs mechanism
the potential $V$ acts only as a source of mass,
the issue of whether the massive modes propagate dynamically 
depends on the structure of these kinetic terms.
In particular,
propagating massive modes can be expected
only if the theory 
\it without \rm
the potential $V$ 
contains the corresponding propagating massless modes.

It is desirable that any propagating modes 
be unitary and ghost free. 
To avoid unitarity issues with higher-derivative propagation,
the kinetic term for the metric excitation $h_{\mu\nu}$
can be taken to emerge as usual
from the Einstein-Hilbert action.
In the absence of the potential $V$,
this is also a ghost-free choice.
For the tensor field,
higher-derivative propagation can be avoided by writing
the general kinetic term $\cl_{\rm K}$ 
as a weighted sum of all scalar densities
formed from quadratic expressions in $T_{\la\mu\nu\cdots}$
that involve two covariant derivatives.
For example, 
one such scalar density is 
\beq
\cl_{\rm K} \sim 
e T_{\la\mu\nu\cdots} D_\al D^\al T^{\la\mu\nu\cdots},
\label{kinetic}
\eeq
where
$e = \sqrt{-g}$ is the vierbein determinant.
In the absence of the potential $V$,
the ghost-free requirement places strong constraints
on allowed forms of $\cl_{\rm K}$
and typically involves gauge symmetry for the tensor.
For a vector field,
for example,
the Maxwell action is the unique ghost-free combination
in the Minkowski spacetime limit.
In any case,
since the massive modes are combinations of
$h_{\mu\nu}$ and $\ta_{\la\mu\nu\cdots}$,
it follows that ghost-free propagation is possible 
only if the kinetic terms 
for these combinations are ghost free. 
Note that the potential $V$ 
may explicitly break the tensor gauge symmetry,
so requiring ghost-free kinetic terms
is by itself insufficient to ensure ghost-free massive modes.

Under the assumption of Lorentz invariance,
the Fierz-Pauli action involving quadratic terms for $h_{\mu\nu}$ 
is the unique ghost-free choice for a free massive spin-2 field
\cite{fp}.
However,
when coupled to a covariantly conserved energy-momentum tensor,
the small-mass limit of a massive spin-2 field 
includes modes that modify the gravitational bending of light
in disagreement with observation 
\cite{vdvz}.
This presents an obstacle to constructing 
a viable theory of massive gravity.
One avenue of investigation that might permit evading 
this obstacle is to relax the assumption of Lorentz invariance. 
Spontaneous Lorentz violation in closed string theory
has been proposed 
\cite{modgrav1}
as a mechanism that might lead to graviton mass terms.
Models involving infrared modifications of gravity 
have also been proposed 
that involve spontaneous Lorentz violation with ghosts
\cite{modgrav2}
and that have explicit Lorentz violations 
\cite{modgrav3}.
Explicit Lorentz violation
is incompatible with Riemann and Riemann-Cartan geometries
but may be compatible with Finsler or other geometries
\cite{akgrav,gyb}
or may be viewed as an approximation to spontaneous violation.
In the present context,
the possibility exists 
that the massive modes from the alternative Higgs mechanism
could evade the Veltman-van Dam-Zakharov constraint
via their origin in spontaneous Lorentz violation 
and their nature as mixtures of gravitational and tensor modes.
For example,
the expansion \rf{Vexpans}
includes quadratic terms involving
contractions of $h_{\mu\nu}$ with the vacuum value $t_{\la\mu\nu}$,
so a model with suitable vacuum values
and incorporating also a ghost-free propagator 
for the corresponding modes 
would describe a type of propagating massive gravity
without conventional Fierz-Pauli terms.

In the alternative Higgs mechanism,
the existence of propagating massive modes involving the metric
can be expected to affect gravitational physics.
Effects can arise directly from the modified graviton propagator
and also from the massive modes acting as sources 
for gravitational interactions.
The latter can be understood by considering
the energy-momentum tensor $T^{\mu\nu}$ of the full theory,
which can be found by variation of the action 
with respect to the metric $g_{\mu\nu}$. 
The contribution $T_V^{\mu\nu}$ to $T^{\mu\nu}$ 
arising from the potential $V$ is
\beq
T_V^{\mu\nu} = 
- g^{\mu\nu} V + 2 V_m^\prime \fr {\de X_m} {\de g_{\mu\nu}} ,
\label{TV}
\eeq
where a sum on $m$ is understood.
For the vacuum solution $X_m = 0$,
which satisfies $V = V_m^\prime = 0$,
$T_V^{\mu\nu}$ remains zero
and the gravitational sourcing is unaffected.
The same is true for excitations 
for which $V$ and $V_m^\prime$ both vanish,
such as the NG modes.
However,
the massive modes arising from a smooth potential 
have nonzero $V$ and $V_m^\prime$
and can therefore act as additional sources
for gravity. 
These contributions can 
lead to a variety of gravitational effects including,
for example,
modifications of the Newton gravitational potential 
in the weak-field limit,
which could have relevance for dark matter,
or cosmological features such as dark energy
\cite{akgrav}.

Note that nonpropagating modes 
can also have similar significant physical effects
on gravitational properties.
This possibility holds for any excitations appearing in $V$,
whether they are physical, ghost, or Lagrange-multiplier modes. 
For example,
any Lagrange-multiplier fields are auxiliary by construction
and so the $\la_m$ excitations are nonpropagating.
Nonetheless,
for linear Lagrange-multiplier potentials
these excitations can contribute to $T_V^{\mu\nu}$ 
even though $V = 0$
because $V_m^\prime$ is nonvanishing.
However,
a theory with a quadratic Lagrange-multiplier potential
has $T_V^{\mu\nu}=0$
and therefore leaves the gravitational sourcing unaffected.

\section{Bumblebee Models}
\label{Bumblebee Models}

In this section,
we focus attention on the special class of theories
in which the role of the tensor $T_{\la\mu\nu\cdots}$
is played by a vector $B_\mu$
that acquires a nonzero vacuum expectation value $b_\mu$.
These theories, 
called bumblebee models,
are among the simplest examples of field theories
with spontaneous Lorentz and diffeomorphism breaking.
In what follows,
bumblebee models are defined,
their properties under local Lorentz and
diffeomorphism transformations are presented,
their mode content is analyzed,
and issues involving 
gauge fixing and alternative mode expansions are considered.

\subsection{Basics}
\label{Basics}

The action $S_B$ for a single bumblebee field $B_\mu$
coupled to gravity and matter can be written as 
\bea
S_B &=& 
\int d^4 x~ \cL_B 
\nonumber\\
&=& 
\int d^4 x~ 
(\cl_g + \cl_{gB} + \cl_{\rm K} + \cl_V + \cl_{\rm M}).
\eea
In Riemann spacetime,
the pure gravitational piece $\cl_g$ 
of the Lagrange density
is usually taken to be the Einstein-Hilbert term
supplemented by the cosmological constant $\La$.
The gravity-bumblebee couplings are described by $\cl_{gB}$,
while $\cl_{\rm K}$ contains the bumblebee kinetic terms 
and any self-interaction terms.
The component $\cl_V$ consists of the potential $V(B_\mu)$,
including terms triggering the spontaneous Lorentz violation.
Finally,
$\cl_{\rm M}$ involves the bumblebee coupling 
to the matter or other sectors in the model.

The forms of $\cL_{gB}$ and $\cL_K$ are complicated
in the general case.
However,
if attention is limited to terms quadratic in $B_\mu$ 
involving no more than two derivatives,
then only five possibilities exist.
The Lagrange density $\cl_B$ can then be written as
\bea
\cl_B &=&
\fr 1 {16\pi G} e (R - 2 \La)
+ \son eB^\mu B^\nu R_{\mu\nu} + \stw e B^\mu B_\mu R
\nonumber\\
&&
- \frac 14 \ton e B^{\mu\nu} B_{\mu\nu}
+ \frac 12 \ttw e D_\mu B_\nu D^\mu B^\nu
\nonumber\\
&&
+ \frac 12 \tth e D_\mu B^\mu D_\nu B^\nu
- eV
+ \cL_{\rm M} ,
\label{bb}
\eea
where $G$ is the Newton gravitational constant
and the field-strength tensor $B_{\mu\nu}$ 
is defined as
\beq
B_\mn = \prt_\mu B_\nu-\prt_\nu B_\mu .
\label{Bfield}
\eeq

The five real parameters 
$\son$, $\stw$, $\ton$, $\ttw$, $\tth$
in Eq.\ \rf{bb}
are not all independent.
Up to surface terms,
which leave unaffected the equations of motion
from the action,
the condition 
\bea
\int d^4x~
(
eB^\mu B^\nu R_{\mu\nu} 
- \frac 12 e B^{\mu\nu} B_{\mu\nu}
\hskip 50pt
\nonumber\\
+ e D_\mu B_\nu D^\mu B^\nu
- e D_\mu B^\mu D_\nu B^\nu
) = 0 
\label{cond}
\eea
is identically satisfied,
so only four of the corresponding five terms in $\cl_B$
are independent.
No term of the form $e B^\mu B_\mu R$
appears in the condition \rf{cond},
so the parameter $\stw$ remains unaffected
while the four parameters $\son$, $\ton$, $\ttw$, $\tth$
become linked.
For example,
the identity \rf{cond} implies that 
the action for the Lagrange density \rf{bb} 
with five nonzero parameters
$\son$, $\stw$, $\ton$, $\ttw$, $\tth$
is equivalent to an action of the same form
but with only four nonzero parameters
$\son^\prime = \son + \tth$, 
$\stw^\prime = \stw$,
$\ton^\prime = \ton - 2 \tth$, 
$\ttw^\prime = \ttw - 2 \tth$, 
while $\tth^\prime = 0$.
Moreover,
other factors may also constrain some of the five parameters.
For example,
certain models of the form $\cl_B$
yield equations of motion that imply  
additional relationships among the parameters.
Also,
certain physical limits
such as the restriction to Minkowski spacetime 
can limit the applicability of the condition \rf{cond}.
Some cases with specific parameter values 
may be of particular interest for 
reasons of physics, geometry, or simplicity,
such as the models with 
$\ton = 1$, $\ttw = \tth = 0$
considered below,
or the model with $\stw = - \son/2$
for which the bumblebee-curvature coupling involves 
the Einstein tensor, 
$\cl_B \supset \son eB^\mu B^\nu G_{\mu\nu}$.
Since the most convenient choice of parameters
depends on the specific model
and on the physics being addressed,
it is useful to maintain the five-parameter form \rf{bb}
for generality.

Following the discussion in 
Sec.\ \ref{Potentials},
we suppose the potential $V$ in Eq.\ \rf{bb}
has no derivative couplings
and is formed from scalar combinations
of the bumblebee field $B_\mu$ 
and the metric $g_{\mu\nu}$.
Only one independent scalar exists.
It can be taken as the bumblebee version of $X$ in Eq.\ \rf{X}:
\beq
X = B_\mu g^{\mu\nu} B_{\nu} \pm b^2,
\label{XB}
\eeq
where $b^2$ is a real nonnegative constant.
The potential $V(X)$ itself can be smooth in $X$,
like the form \rf{VS}, 
or it can involve Lagrange multipliers
like the form \rf{VL} or \rf{VQ}.
In any case,
the vacuum is determined by the single condition
\beq
X = B_\mu g^{\mu\nu} B_\nu \pm b^2 = 0  .
\label{Bsquare}
\eeq
In the vacuum,
the potential vanishes, $V(X)=0$,
and the fields $B_\mu$, $g_{\mu\nu}$ 
acquire vacuum values
\beq
B_\mu \to \vev{B_\mu} = b_\mu , \quad
g_{\mu\nu} \to \vev{g_{\mu\nu}}.
\label{vevs}
\eeq
The nonzero value $b_\mu$,
which obeys $b_\mu \vev{g^{\mu\nu}}b_\nu = \mp b^2$,
spontaneously breaks 
both Lorentz and diffeomorphism symmetry.
Note that the choice of the potential $V$
can also have implications for the parameters in Eq.\ \rf{bb}.
For example,
if the potential takes 
the quadratic Lagrange-multiplier form \rf{VQ},
then $\stw$ can be taken as zero 
because a nonzero value merely acts to rescale $G$ and $\La$.
However,
a nonzero value of $\stw$ can have nontrivial consequences 
for models with other potentials,
such as the smooth quadratic form \rf{VS}.

For generic choices of parameters,
the Lagrange density \rf{bb} 
is unitary because no more than two derivatives appear.
However,
as discussed in Sec.\ \ref{Massive modes},
the indefinite metric and generic absence of gauge invariance 
typically imply the presence of ghosts 
and corresponding negative-energy problems,
which can tightly constrain the viability of various models
\cite{ems}.
If the gravitational couplings and the potential $V$
are disregarded,
there is a unique set of parameters
ensuring the absence of ghosts:
$\ton = 1$, $\ttw = \tth = 0$.
With this choice,
the kinetic term for the bumblebee becomes
the Maxwell action,
in which the usual U(1) gauge invariance excludes ghosts.
When the gravitational terms and couplings are included,
this gauge invariance is maintained 
for the parameter choice 
$\ton = 1$, $\ttw = \tth = \son = \stw  = 0$.
The further inclusion of the potential $V$
breaks the U(1) gauge symmetry,
but the form of the kinetic term
ensures a remnant constraint 
on the equations of motion
arising from the identity
\beq
D_\mu D_\nu B^{\mu\nu} \equiv 0.
\label{remconst}
\eeq
The action for these bumblebee models,
introduced in Ref.\ \cite{ks},
is therefore of particular interest.
The corresponding Lagrange density 
\bea
\cl_{\rm \ks} &=&
\fr 1 {16\pi G} e (R - 2 \La)
- \frac 14 e B^{\mu\nu} B_{\mu\nu}
- eV + \cL_{\rm M} ,
\qquad
\label{ksbb}
\eea
is investigated in more detail in Sec.\ \ref{Examples}.
The reader is warned 
that some confusion about the relationship between 
these models and ones with nonzero $\ttw$ and $\tth$
exists in the literature.
In particular,
results for the models \rf{ksbb}
can differ from those obtained 
in models with nonzero $\ttw$, $\tth$
via straightforward adoption 
of the limit $\ttw, \tth \to 0$,
due to the emergence of the remnant constraint \rf{remconst}.

The discussion above defines bumblebee models 
on the spacetime manifold,
without introducing a local Lorentz basis.
In this approach,
the Lorentz NG modes are hidden 
within the bumblebee field $B_\mu$.
Adopting instead a vierbein formulation
reveals explicitly the local Lorentz properties of the models,
and it also provides a natural way to incorporate 
spinor fields in the matter Lagrange density $\cl_{\rm M}$.
The vierbein $\vb \mu a$
converts tensors expressed in a local basis 
to ones on the spacetime manifold.
The spacetime metric $g_{\mu\nu}$ is related
to the Minkowski metric $\et_{ab}$ in the local frame as
\beq
g_{\mu\nu} = \vb \mu a \vb \nu b  \et_{ab} ,
\label{gvb}
\eeq
while the bumblebee spacetime vector $B_\mu$ 
is related to the local bumblebee vector $B_a$ as 
\beq
B_\mu = \vb \mu a B_a .
\label{Bvb}
\eeq
A complete treatment in the vierbein formalism 
involves also introducing the spin connection $\nsc \mu a b$,
which appears in covariant derivatives acting on local quantities.
In Riemann-Cartan geometry,
where the spacetime has both curvature and torsion,
the spin connection represents degrees of freedom
independent of the vierbein.
However,
experimental constraints on torsion are tight
\cite{torsion}.
In this work,
we restrict attention to Riemann geometry,
for which the torsion vanishes 
and the spin connection is fixed in terms of the vierbein.
It therefore suffices for our purposes to  
consider the vierbein degrees of freedom 
and their role relative to the gravitational and NG modes.
Bumblebee models in the more general context 
of Riemann-Cartan spacetime with nonzero torsion
are investigated in Refs.\ \cite{akgrav,bk}.

There is a substantial literature concerning
theories of vacuum-valued vectors coupled to gravity.
The five-parameter Lagrange density \rf{bb} 
excluding the potential $V$ and the cosmological constant
was investigated by Will and Nordtvedt
in the context of vector-tensor models of gravity 
\cite{wn}.
Kosteleck\'y and Samuel (\ks)
\cite{ks}
introduced the potential $V$
triggering spontaneous Lorentz violation 
and studied both the smooth quadratic potential 
\rf{VS}
and the linear Lagrange-multiplier case
\rf{VL}
for the class of models given by Eq.\ \rf{ksbb}.
The presence of the potential 
introduces a variety of qualitatively new effects,
including 
the necessary breaking of U(1) gauge invariance
\cite{akgrav},
the existence of massless NG modes and massive modes 
\cite{bk},
and potentially observable novel effects 
for post-newtonian physics
\cite{bak06}
and for the matter sector
\cite{kleh}.
The potential $V$ also leads 
to a candidate alternative description of the photon
\cite{bk}
and the graviton
\cite{akgrav,kpgr}.
In Minkowski spacetime,
more general potentials $V$ of hypergeometric form 
are known to satisfy the one-loop exact renormalization group 
\cite{baak}.
Models of the form \rf{bb}
with $\son = \stw = 0$,
a unit timelike $b_\mu$,
a linear Lagrange-multiplier potential $V_L$,
and an additional fourth-order interaction for $B_\mu$
have been studied in some detail
as possible unconventional theories of gravity
\cite{bb1,bb2}.
Other works involving bumblebee models include
Refs.\ \cite{ems,cli,bmg,gjw,cfn,bb3}.

An aspect of bumblebee models of particular interest
is the appearance of massless propagating vector modes.
This feature suggests the prospect of an alternative 
to the usual description of photons via U(1) gauge theory.
The central idea is to identify the photon modes 
with the NG modes arising from spontaneous Lorentz violation.
Early discussions along these lines
centered on reinterpretating the photon or electron 
in the context of special relativity
without physical Lorentz violation
\cite{dhfb,yn}.
More recently,
the Lorentz NG modes in certain bumblebee models 
with physical Lorentz violation
have been shown to obey the Einstein-Maxwell equations
in Riemann spacetime in axial gauge
\cite{bk}.
These models are further considered below, 
with the discussion initiated
in Sec.\ \ref{Bumblebee electrodynamics}.
One motivation of the present work
is to investigate the role of massive modes
and Lagrange-multiplier fields in this context.
In particular,
it is of interest to investigate possible modifications 
to electrodynamics and gravity,
with an eye to novel phenomenological applications 
of spontaneous Lorentz and diffeomorphism breaking.

\subsection{Transformations}
\label{Transformations}

In considering theories 
with violations of Lorentz and diffeomorphism symmetry,
it is important to distinguish
between {\it observer} and {\it particle} transformations
\cite{ck}.
Under either an observer general coordinate transformation
or an observer local Lorentz transformation,
geometric quantities
such as scalars, vectors, tensors, and their derivatives 
remain unchanged,
but the coordinate basis defining their components transforms.
In contrast,
particle transformations can change geometric quantities,
independently of any coordinate system and basis. 

In theories without spacetime-symmetry breaking,
the component forms of the transformation laws 
for particle and observer transformations are inversely related.
For example,
under infinitesimal particle diffeomorphisms
described by four infinitesimal dispacements $\xi^\mu$,
the components of the bumblebee field transform
according to the Lie derivative as
\bea
B_\mu 
&\rightarrow &
 B_\mu - (\prt_\mu \xi^\nu) B_\nu
- \xi^\nu \prt_\nu B_\mu ,
\nonumber\\
B^\mu 
&\rightarrow &
B^\mu + (\prt_\nu \xi^\mu) B^\nu
- \xi^\nu \prt_\nu B^\mu,
\label{Bdiff}
\eea
while the metric transforms as
\bea
g_{\mu\nu} 
&\rightarrow &
g_{\mu\nu} - (\prt_\mu \xi^\al) g_{\al\nu}
- (\prt_\nu \xi^\al) g_{\mu\al} - \xi^\al \prt_\al g_{\mu\nu} ,
\nonumber\\
g^{\mu\nu} 
&\rightarrow &
g^{\mu\nu} + (\prt_\al \xi^\mu) g^{\al\nu}
+ (\prt_\al \xi^\nu) g^{\mu\al} - \xi^\al \prt_\al g^{\mu\nu} .
\qquad
\label{gdiff}
\eea
Under infinitesimal observer general coordinate transformations,
which are the observer equivalent of diffeomorphisms,
the formulae for the transformations
of the bumblebee and metric components
take the same mathematical form
up to a possible sign change 
in the arbitrary parameter $\xi^\mu$,
even though these transformations 
are only the result of a change of coordinates.
Similarly,
under infinitesimal particle local Lorentz transformations
with six parameters $\ep_{ab} = -\ep_{ba}$ 
related to the local Lorentz group element by
$\La_a^{\pt{a} b} \approx \de_a^{\pt{a} b} + \ep_a^{\pt{a} b}$,
the local components of the bumblebee field transform as
\beq
B_a \rightarrow B_a + \ep_a^{\pt{a} b} B_b ,
\label{BLT}
\eeq
with the formula for $B^a$ following
from this by raising indices with $\et^{ab}$.
Under observer local Lorentz transformations,
the corresponding transformation formulae 
again take the same mathematical form.
However,
any form of physical Lorentz and diffeomorphism breaking 
destroys the details of these equivalences.

A fundamental premise in classical physics
is that the properties of a physical system
are independent of the presence of a noninteracting observer.
The observer is free to select a coordinate system
to describe the physics of the system,
but the physics cannot depend on this choice.
In particular,
this must remain true
even when Lorentz and diffeomorphism symmetry are broken,
whether explicitly or spontaneously.
Viable candidate theories with spacetime-symmetry breaking
must therefore be invariant
under observer transformations,
which are merely changes of coordinate system.
For example,
the SME is formulated as a general effective field theory 
that is invariant 
under observer general coordinate transformations
and under observer local Lorentz transformations
\cite{akgrav,kp,ck}.

In contrast,
a theory with physical Lorentz and diffeomorphism breaking
cannot by definition remain fully invariant 
under the corresponding particle transformations.
For example,
if the breaking is explicit,
the action of the theory changes under particle transformations.
If instead the breaking is spontaneous,
the action remains invariant
and the equations of motion transform covariantly,
as usual.
However,
the vacuum solution to the equations of motion
necessarily contains quantities with spacetime indices
that are unchanged by particle transformations.
These vacuum values and the excitations around them
can lead to physical effects revealing the symmetry breaking.

In bumblebee models,
the relevant spacetime vacuum values are those 
of the bumblebee and metric fields,
denoted $\vev{B_\mu}$ and $\vev{g_{\mu\nu}}$.
These are unaffected by particle diffeomorphisms,
and their nonzero components thereby reveal 
the broken particle diffeomorphisms.
In contrast,
$\vev{B_\mu}$ and $\vev{g_{\mu\nu}}$
both transform as usual
under observer general coordinate transformations,
which therefore are unbroken.
Analogous results hold for the local-frame vacuum value $\vev{B_a}$ 
of the local bumblebee field.
The components $\vev{B_a}$ remain unaffected
by particle local Lorentz transformations,
with the invariance of the nonzero components
resulting from the breaking of local Lorentz generators,
while the components $\vev{B_a}$ transform 
under observer local Lorentz transformations 
in the usual way.
Similarly,
the vacuum value $\vev{\vb \mu a}$ of the vierbein
is unchanged by both particle diffeomorphisms 
and particle Lorentz transformations,
but it transforms as a vector under the
corresponding observer transformations.

By virtue of the relation \rf{Bvb}
between the spacetime and local bumblebee fields,
it follows that spontaneous local Lorentz violation
is necessarily accompanied by spontaneous diffeomorphism violation
and vice versa
\cite{bk}.
The point is that the vacuum value of the vierbein is nonzero,
so the existence of a nonzero $\vev{B_a}$ 
spontaneously breaking local Lorentz symmetry
also implies the existence of a nonzero $\vev{B_\mu}$ 
spontaneously breaking diffeomorphism symmetry,
and vice versa.
Note, however, 
that this result fails for explicit violation,
where the analogue of the relation \rf{Bvb} is absent.
In general,
explicit local Lorentz violation occurs
when a nonzero quantity 
$t_{abc\ldots}$
is externally prescribed in the local frame,
but the corresponding spacetime quantity 
$t_{\la\mu\nu\ldots}\equiv 
\vb \la a \vb \mu b \vb \nu c \cdots
t_{abc\ldots}$
is defined using the full vierbein
and hence remains invariant under diffeomorphisms.
Similarly,
explicit diffeomorphism violation occurs
when a nonzero quantity $t_{\la\mu\nu\ldots}$
is externally prescribed
on the spacetime manifold,
but the corresponding local quantity 
defined via the inverse vierbein
remains invariant under local Lorentz transformations.
 
\subsection{Mode expansions}
\label{Mode expansions}

To study the content and behavior of the modes,
the fields can be expanded as infinitesimal excitations
about their vacuum values.
At the level of the Lagrange density,
it suffices for most purposes to keep only terms to second order 
in the field excitations,
which linearizes the equations of motion.

Assuming for simplicity a Minkowski background
$\vev{g_{\mu\nu}} = \et_{\mu\nu}$,
we expand the metric and its inverse as
\beq
g_{\mu\nu} = \et_{\mu\nu} + h_{\mu\nu},
\quad
g^{\mu\nu} \approx \et^{\mu\nu} - h^{\mu\nu} .
\label{gh}
\eeq
For the local bumblebee vector,
we write
\beq
B_a = b_a + \be_a ,
\label{Bbe}
\eeq
where the vacuum value $\vev{B_a}$ is denoted $b_a$
and assumed constant,
and 
the infinitesimal excitations are denoted $\be_a$.
The vacuum condition \rf{Bsquare} implies
\beq
b_a \et^{ab} b_b = \mp b^2.
\eeq

In a Minkowski background,
the vierbein vacuum value can be chosen to be 
$\vev{\vb \mu a} = \de_\mu^{\pt{\mu}a}$
in cartesian coordinates.
The expansions of the vierbein and its inverse are 
\bea
\vb \mu a 
&\approx &
\de_\mu^{\pt{\mu}a} + \half h_\mu^{\pt{\mu}a}
+ \ch_\mu^{\pt{\mu}a} ,
\nonumber\\
\ivb \mu a 
&\approx &
\de^\mu_{\pt{\mu}a} - \half h^\mu_{\pt{\mu}a}
+ \ch^\mu_{\pt{\mu}a} ,
\label{vhch}
\eea
where $h_{\mu a}$ and $\ch_{\mu a}$ are,
respectively,
the symmetric and antisymmetric components of the vierbein.
The covariant and contravariant components of the bumblebee field 
follow from Eq.\ \rf{Bvb} as
\bea
B_\mu 
&\approx &
\de_\mu^{\pt{\mu}a} (b_a + \be_a)
+ ( \half h_\mu^{\pt{\mu}a} + \ch_\mu^{\pt{\mu}a}) b_a ,
\nonumber\\
B^\mu 
&\approx &
\de^\mu_{\pt{\mu}a} (b^a + \be^a)
+ (-\half h^\mu_{\pt{\mu}a} + \ch^\mu_{\pt{\mu}a}) b^a .
\label{Bvbup}
\eea

Since the local and spacetime background metrics 
are both Minkowski
and since the excitations are infinitesimal,
the distinction between the Latin local indices
and the Greek spacetime indices can be dropped.
We adopt Greek indices for most purposes that follow, 
raising and lowering indices on purely first-order quantities
with $\et^{\mu\nu}$ and $\et_{\mu\nu}$.
The vierbein expansions \rf{vhch} can then be rewritten as
\bea
\lvb \mu \nu 
&\approx &
\et_{\nu\si} \vb \mu \si
\approx ~
\et_{\mu\nu} + \half h_{\mu\nu} + \ch_{\mu\nu} ,
\nonumber\\
\uvb \mu \nu 
&\approx &
\et^{\nu\si} \ivb \mu \si
\approx ~
\et^{\mu\nu} - \half h^{\mu\nu} + \ch^{\mu\nu} .
\label{vhch2}
\eea
The vacuum value for the bumblebee vector becomes
\beq
\vev{B_\mu} 
= \vev{\vb \mu a}  b_a 
= \de_\mu^{\pt{\mu}a} b_a \equiv b_\mu.
\label{bbvac}
\eeq

It is convenient to decompose the local vector
excitations $\be_\mu \equiv \de_\mu^{\pt{\mu}a} \be_a$
into transverse
and longitudinal pieces with respect to $b_\mu$.
Excluding for simplicity
the case of lightlight $b_\mu$,
we write
\beq
\be_\mu = \be^{\rm t}_\mu + \be \hat b_\mu ,
\quad \be^{\rm t}_\mu b^\mu = 0,
\label{beta}
\eeq
where $\hat b_\mu = b_\mu/\sqrt{b^2}$
is a vector along the direction of $b_\mu$ obeying
$\hat b^\mu \hat b_\mu = \mp 1$.
Using this decomposition,
the bumblebee mode expansions \rf{Bvbup} become
\bea
B_\mu 
&\approx & 
b_\mu
+ (\half h_{\mu\nu} + \ch_{\mu\nu}) b^\nu
+ \be^{\rm t}_\mu + \be \hat b_\mu ,
\nonumber\\
B^\mu 
&\approx & 
b^\mu
+ (- \half h^{\mu\nu} + \ch^{\mu\nu}) b_\nu
+ \be^{{\rm t}\mu} + \be \hat b^\mu .
\label{Bvb2}
\eea

It is instructive to count degrees of freedom
in the expressions \rf{vhch2} and \rf{Bvb2}.
On the left-hand side, 
the vierbein has 16 components 
and the bumblebee field 4, for a total of 20.
On the right-hand side,
the symmetric metric component $h_{\mu\nu}$ has 10,
the antisymmetric component $\ch_{\mu\nu}$ has 6,
the transverse bumblebee excitation $\be^{\rm t}_\mu$ has 3,
and the longitudinal excitation $\be$ has 1,
producing the required matching total of 20.
Of these 20 degrees of freedom,
6 are metric modes,
4 are bumblebee modes,
while 6 are associated with local Lorentz transformations
and 4 with diffeomorphisms.
The explicit transformations of all the field components
under particle diffeomorphisms 
and local Lorentz transformations 
can be deduced from the full-field expressions
and from the invariance of the vacuum values. 
A list of formulae 
for both particle and observer transformations 
is provided in the Appendix.

Among all the excitations,
only the longitudinal component $\be$ 
of the bumblebee field is invariant 
under both particle diffeomorphisms and local Lorentz transformations.
It is therefore a physical degree of freedom
in any gauge.
Moreover,
using Eqs.\ \rf{Bsquare} and \rf{Bvb2} 
reveals that exciting the $\be$ mode alone 
produces a nonzero value of $X$,
given at first order by 
\beq
X = B_\mu g^{\mu\nu} B_\nu \pm b^2 \approx 2 (b^\mu \hat b_\mu) \be .
\label{Bsquare2}
\eeq
As a result,
the excitation $\be$ is associated with a 
nonminimal value of the potential,
and it therefore cannot be an NG mode.
In fact,
for the case of a smooth quadratic potential,
$\be$ is a massive mode.
In contrast,
in a theory with a Lagrange-multiplier potential
where the constraint $X = 0$ is enforced 
as an equation of motion,
the massive mode $\be$ identically vanishes.

\subsection{Gauge fixing and NG modes}
\label{Gauge fixing and NG modes}

Since the spacetime-symmetry breaking in bumblebee models 
is spontaneous,
the four diffeomorphisms parametrized by $\xi_\mu$ 
and the six local Lorentz transformations
parametrized by $\ep_{\mu\nu}$
leave invariant the bumblebee action \rf{bb}
and transform covariantly the equations of motion.
Fixing the gauge freedom therefore requires 
10 gauge conditions.

For the diffeomorphism freedom,
a choice common in the literature is the harmonic gauge
\beq
\prt_\mu \ol h^{\mu\nu}  = 0 ,
\label{harm}
\eeq
where
$\ol h^{\mu\nu} =  h^{\mu\nu} - \half \et_{\mu\nu} h$
and $\ol h = - h \equiv - h^\mu_{\pt{\mu}\mu}$.
In the harmonic gauge,
the Einstein tensor becomes 
$G_{\mu\nu} \approx -\half \square \ol h_{\mu\nu}$ at linear order.
An alternative choice for the diffeomorphism degrees of freedom
is the axial gravitational gauge 
\beq
h_{\mu\nu} b^\nu = 0.
\label{haxial}
\eeq
Both these gauge choices represent four conditions.
To fix the local Lorentz freedom,
it is common to eliminate 
all six antisymmetric vierbein components
by imposing the six conditions
\beq
\ch_{\mu\nu} = 0.
\label{chigauge}
\eeq
Other choices are possible here too. 
Consider,
for example,
the decomposition of $\ch_{\mu\nu}$
in terms of projections along $b_\mu$,
\bea
\ch_{\mu\nu} &=& 
\ch^{tt}_{\mu\nu} 
\mp \ch_\mu^t \hat b_\nu 
\pm \ch_\nu^t \hat b_\mu ,
\nonumber\\
\ch^{tt}_{\mu\nu} b^\nu 
&=& 
\ch^t_\nu b^\nu = 0,
\label{chidec}
\eea
which is the analogue of Eq.\ \rf{beta} for $\be_\mu$.
The components $\ch^{tt}_{\mu\nu}$
and $\ch_\mu^t \equiv \ch_{\mu\nu} \hat b^\nu$
each contain three degrees of freedom.
Inspection of the transformation laws shows that 
an alternative to fixing the local Lorentz gauge 
via Eq.\ \rf{chigauge}
is the set of six conditions
\beq
\ch^{tt}_{\mu\nu} = \be^t_\mu = 0.
\label{btgauge}
\eeq
Note,
however,
that the combination 
$(\ch_{\mu\nu} b^\nu + \be^{\rm t}_\mu)$
appearing in $B_\mu$ 
in Eq.\ \rf{Bvb2}
is invariant 
and therefore cannot be gauged to zero.
Evidently,
the associated local Lorentz degrees of freedom 
must remain somewhere in the theory.

The bumblebee vacuum value $\vev{B_\mu} = b_\mu$
breaks one of the four diffeomorphism symmetries.
The broken generator is associated with
the projected component $\xi_\nu b^\nu$ 
of the parameter $\xi_\mu$. 
Analogously,
the vacuum value $\vev{B_a} = b_a$ breaks
three of the six local Lorentz symmetries.
The broken generators are associated with the components
$\ep_{ab} b^b \approx \ep_{\mu\nu} b^\nu$ 
of $\ep_{\mu\nu}$ projected along $b^\nu$.
In each case,
the unbroken generators are associated with the 
orthogonal complements to the projections.

Since the vacuum breaks one particle diffeomorphism
and three local Lorentz transformations,
four NG modes appear.
A useful general procedure to identify these modes
is first to make virtual particle transformations 
using the broken generators
acting on the appropriate vacuum values 
for the fields in $V$,
and then to promote the corresponding parameters 
$\ep_{\mu\nu}$ and $\xi_\mu$ to field excitations
\cite{bk}:
\beq
\xi_\mu \to \Xi_\mu,
\quad
\ep_{\mu\nu} \to \cE_{\mu\nu} = - \cE_{\nu\mu}.
\label{promotion}
\eeq
The properties of $\Xi_\mu$ and $\cE_{\mu\nu}$ 
under various particle and observer transformations
are given in the Appendix.
The projections 
$\xi_\nu b^\nu$ and $\ep_{\mu\nu} b^\nu$ 
associated with the broken generators
determine the NG modes,
which are therefore 
$\Xi_\nu b^\nu$ and $\cE_{\mu\nu} b^\nu$.

We can follow this procedure to elucidate 
the relationship between 
the NG modes and the component fields
in the decomposition \rf{Bvb2} of $B_\mu$.
Consider first the diffeomorphism NG mode,
which is generated by a virtual transformation 
acting on the spacetime vacuum value $\vev{B_\mu}$ 
and involving the broken diffeomorphism generator.
The relevant transformation is given in Eq.\ \rf{pdiffs},
and it yields 
\beq
\vev {B_\mu} \to b_\mu - (\prt_\mu \Xi_\nu) b^\nu . 
\eeq
Comparison of this result to the form of Eq.\ \rf{Bvb2}
reveals that the four vierbein combinations
$(\half h_{\mu\nu} + \ch_{\mu\nu})b^\nu$
contain the diffeomorphism NG mode.
This agrees with the result obtained by combining 
virtual diffeomorphisms on the vacuum values
of the component fields in Eq.\ \rf{Bvb2}:
\bea
\vev{h_{\mu\nu} b^\nu}
&\to & 
-(\prt_\mu \Xi_\nu + \prt_\nu 
\Xi_\mu) b^\nu ,
\nonumber\\ 
\vev{\ch_{\mu\nu} b^\nu}
&\to & 
- \half (\prt_\mu \Xi_\nu - \prt_\nu \Xi_\mu) b^\nu ,
\nonumber\\ 
\vev{\be^t_\mu} 
&\to &
0, 
\quad
\vev{\be} \to 0.
\label{XiEpredef}
\eea
Note also that 
the combinations $(\half h_{\mu\nu} + \ch_{\mu\nu})b^\nu$
maintain the potential minimum $V=0$,
as is expected for an NG mode.

In contrast,
the three Lorentz NG modes $\cE_{\mu\nu} b^\nu$
are generated by virtual transformations
acting on the local-frame vacuum value $\vev{B_a}$ 
and involving the broken local Lorentz generators:
\beq
\vev {B_a} \to b_a + \cE_{ab} b^b ,
\label{LLvirttr}
\eeq
Comparison with Eqs.\ \rf{Bbe} and \rf{beta}
shows that the transverse field $\be^t_\mu$
contains the three Lorentz NG modes $\cE_{\mu\nu} b^\nu$.
Note that there are exactly three degrees of freedom 
in $\be^t_\mu$,
all of which maintain the potential minimum $V=0$.
This result can be used to connect
the component fields in the decomposition of $B_\mu$
with the three Lorentz NG modes $\cE_{\mu\nu} b^\nu$.
The vacuum value $\vev{B_\mu}$ itself is invariant 
under local Lorentz transformations,
so another component field must also contain
the Lorentz NG modes.
Performing virtual local Lorentz transformations
on the vacuum values of the component fields gives
\bea
\vev{h_{\mu\nu} b^\nu} 
&\to & 
0, 
\quad \vev{\ch_{\mu\nu} b^\nu}
\to 
-\cE_{\mu\nu} b^\nu ,
\nonumber\\ 
\vev{\be^t_\mu} 
&\to &
\cE_{\mu\nu} b^\nu ,
\quad
\vev{\be} \to 0,
\label{vllt}
\eea
which shows that the three combinations
$\vev{\ch_{\mu\nu} b^\nu}$
also contain the three Lorentz NG modes.

Combining the results \rf{XiEpredef} and \rf{vllt}
reveals the following NG mode content 
for the component field combinations
in the decomposition \rf{Bvb2} of $B_\mu$:
\bea
(-\half h_{\mu\nu} + \ch_{\mu\nu}) b^\nu
&= & 
- (\prt_\mu \Xi_\nu) b^\nu - \cE_{\mu\nu} b^\nu ,
\nonumber\\ 
\be^t_\mu
&= &
\cE_{\mu\nu} b^\nu ,
\nonumber\\ 
\be &\supset& 0.
\label{combined}
\eea
We see that the four combinations
$(\half h_{\mu\nu} + \ch_{\mu\nu})b^\nu$
contain a mixture of the diffeomorphism NG mode
and the three Lorentz NG modes,
while the three combinations $\be^t_\mu$
contain only the three Lorentz NG modes.

By fixing the 10 gauge freedoms in the theory, 
the above results can be used to determine 
the physical content of the field $B_\mu$.
Suppose first we adopt 
the 10 conditions \rf{haxial} and \rf{chigauge}.
Then,
the bumblebee field becomes
\bea
B_\mu 
&\approx &
b_\mu + \be^t_\mu + \be \hat b_\mu 
\nonumber\\
&=&
b_\mu + \cE_{\mu\nu} b^\nu + \be \hat b_\mu ,
\label{bgauge1}
\eea
where the result \rf{combined} has been used. 
Note that the fixed gauge means that fields
with different transformation properties 
can appear on the left- and right-hand sides.
In this gauge,
the four components of $B_\mu$
are decomposed into
three Lorentz NG modes associated with $\be^t_\mu$
and the one massive mode $\be$.
The diffeomorphism NG mode is absent.
It obeys
\beq
(\prt_\mu \Xi_\nu)b^\nu = -(\prt_\nu \Xi_\mu) b^\nu
\label{diffeocond1}
\eeq
and hence
$b^\mu(\prt_\mu \Xi_\nu)b^\nu = 0$,
and it is locked to the Lorentz NG modes via the equation 
\beq
(\prt_\mu \Xi_\nu)b^\nu = - \cE_{\mu\nu} b^\nu.
\label{diffeocond2}
\eeq
For the diffeomorphism NG mode,
this gauge is evidently analogous to the unitary gauge
in a nonabelian gauge theory. 

An alternative gauge choice could be to impose
the 10 conditions \rf{haxial} and \rf{btgauge} instead.
This gives
\bea
B_\mu 
&\approx &
b_\mu + \ch^t_\mu + \be \hat b_\mu 
\nonumber\\
&=&
b_\mu + \cE_{\mu\nu} b^\nu + \be \hat b_\mu ,
\label{bgauge2}
\eea
where in this gauge
the Lorentz NG modes $\cE_{\mu\nu} b^\nu$
are identified with $\ch^t_\mu$
instead of $\be^t_\mu$.
One way to understand this identification  
is to perform a local Lorentz transformation
with parameter $\ep_{\mu\nu} = -\cE_{\mu\nu}$
on the first result in Eq.\ \rf{bgauge1}.
This changes the value of $\be^t_\mu$ 
from $\cE_{\mu\nu} b^\nu$ to zero,
while simultaneously converting
$\ch^t_\mu$ from zero
to $\ch^t_\mu =\cE_{\mu\nu} b^\nu$.
In this gauge,
the explicit decomposition \rf{bgauge2} of $B_\mu$ 
in terms of NG modes is the same as that in Eq.\ \rf{bgauge1},
but the three Lorentz NG modes 
are now associated with the components $\ch^t_\mu$
of the vierbein.
The diffeomorphism mode remains absent.
It still obeys the condition \rf{diffeocond1}
and is locked to the Lorentz NG modes
by Eq.\ \rf{diffeocond2}.

Partial gauge conditions can also be imposed.
For example,
suppose the choice \rf{btgauge} 
is made for the local Lorentz gauge,
while the diffeomorphism gauge remains unfixed. 
Then,
the four degrees of freedom in
the combination $(\half h_{\mu\nu} + \ch_{\mu\nu})b^\nu$
consist of the three Lorentz NG modes
and the diffeomorphism NG mode,
and all the NG modes are contained in the vierbein
\cite{bk}. 
The bumblebee field can be written as 
\beq
B_\mu 
\approx 
b_\mu 
-(\prt_\mu \Xi_\nu) b^\nu + \cE_{\mu\nu} b^\nu 
+ \be \hat b_\mu .
\label{Bvbredef}
\eeq
Note that only one projection of $\Xi_\nu$ appears,
even though four diffeomorphism choices remain.
The corresponding expression for $B^\mu$ 
includes additional metric contributions 
and is given by
\beq
B^\mu \approx b^\mu + (\prt_\nu \Xi^\mu) b^\nu 
+ \cE^{\mu\nu} b_\nu 
+ \be \hat b^\mu .
\label{upBvbredef}
\eeq
This equation contains contributions 
from all four fields $\Xi_\mu$.
However,
if the diffeomorphism excitations 
are restricted only to the one for the broken generator,
for which $\Xi_\mu$ obeys $b_\mu \Xi_\nu = b_\nu \Xi_\mu$,
then $B^\mu$ also reduces to an expression 
depending only on the diffeomorphism NG mode $\Xi_\nu b^\nu$.
Related results are obtained in Ref.\ \cite{bk},
which investigates the fate of the NG modes 
using a decomposition of the vierbein 
into transverse and longitudinal components along $b_\mu$.
This decomposition leads to the same relations as 
Eqs.\ \rf{Bvbredef} and \rf{upBvbredef}
when the condition $b_\mu \Xi_\nu = b_\nu \Xi_\mu$ is applied.

Even without gauge fixing,
the fields $\Xi_\mu$ cancel at linear order
in both $G_{\mu\nu}$ and $B_{\mu\nu}$.
As a result, 
propagating diffeomorphism NG modes cannot appear.
This is a special case of a more general result.
By virtue of their origin as virtual particle transformations,
the diffeomorphism NG modes appear as certain components
of representation-irreducible fields with nonzero vacuum values.
However,
diffeomorphism invariance ensures 
these modes enter in the metric and bumblebee fields
in combinations that cancel in a diffeomorphism-invariant action,
including the general bumblebee action \rf{bb}.
In contrast,
the Lorentz NG modes do play a role in the bumblebee action. 
They can be identified with a massless vector field
$A_\mu \equiv \cE_{\mu\nu} b^\nu$
in the fixed axial gauge $A_\mu b^\mu \approx 0$.
The fully gauge-fixed expression \rf{bgauge1} 
for $B_\mu$ then becomes
\beq
B_\mu \approx b_\mu + A_\mu + \be \hat b_\mu ,
\label{bba}
\eeq
where the transverse components 
have the form of photon fields 
in the axial gauge,
and the longitudinal mode is the massive mode $\be$.

\subsection{Alternative expansions}
\label{Alternative expansions}

The analysis in the previous subsections
is based on the local-frame expansion \rf{Bbe}
of the bumblebee field $B_a$.
However, 
other mode expansions are possible,
including ones in which the vierbein makes no explicit appearance
and the local Lorentz transformations are no longer manifest. 
This subsection offers a few comments
on two alternative expansions used in some of the literature.

The first alternative mode expansion
is specified using the covariant spacetime components 
of the bumblebee field
\cite{bk,kp},
\beq
B_\mu = b_\mu + \cE_\mu ,
\label{epdown}
\eeq
where $\cE_\mu$ represents the excitations
of the spacetime bumblebee field 
around the vacuum value $b_\mu$.
It follows that the contravariant components are 
\beq
B^\mu \approx b^\mu + \cE^\mu - h^{\mu\nu} b_\nu .
\label{epup}
\eeq
These fields are linked to the vierbein 
and the local-frame fluctuations $\be_\mu$ via 
\beq
\cE_\mu =
(\half h_{\mu\nu} + \ch_{\mu\nu}) b^\nu
+ \be^{\rm t}_\mu + \be \hat b_\mu .
\label{redef1}
\eeq
The fields $\cE_\mu$ are 
scalars under particle local Lorentz transformations,
but transform under particle diffeomorphisms as
\bea
\cE_\mu &\rightarrow& \cE_\mu - (\prt_\mu \xi_\al) b^\al , 
\nonumber\\
\cE^\mu \equiv \et^{\mu\nu} \cE_\nu &\rightarrow& \cE^\mu - 
\et^{\mu\nu} (\prt_\nu \xi_\al) b^\al .
\label{epdiff}
\eea
Note that this usage of $\cE^\mu$
differs from that in Ref.\ \cite{bk}.

The second alternative expansion 
\cite{bak06,bb1,cli,bmg,gjw}
starts instead with the contravariant bumblebee components
\beq
B^\mu = b^\mu + \cE^{\prime\mu} ,
\label{epprimeup}
\eeq
where a prime is used to distinguish the excitations
$\cE^{\prime\mu}$ from the previous case.
The corresponding covariant components are then 
\beq
B_\mu \approx b_\mu + \cE^\prime_\mu + h_{\mu\nu} b^\nu .
\label{epprimedown}
\eeq
The field redefinitions connecting these to the vierbein
and previous case are
\bea
\cE^{\prime\mu} &=&
(- \half h^{\mu\nu} + \ch^{\mu\nu}) b_\nu
+ \be_{\rm t}^\mu + \be \hat b^\mu
\nonumber\\ 
&=& \et^{\mu\nu} \cE_\mu - h^{\mu\nu} b_\nu .
\label{redef2}
\eea
The fields $\cE^{\prime\mu}$ 
are scalars under particle local Lorentz transformations
but transform under particle diffeomorphisms as 
\bea
\cE^{\prime\mu} 
&\rightarrow& \cE^{\prime\mu} + (\prt_\al \xi^\mu) b^\al ,
\nonumber\\ 
\cE^\prime_\mu \equiv \et_{\mu\nu} \cE^{\prime\mu}
&\rightarrow& \cE^\prime_\mu + (\prt_\al \xi_\mu) b^\al .
\label{epprimediff}
\eea
Note that the fields $\cE_\mu$ and $\cE^\prime_\mu$
have different transformation properties
and contain different mixes of the bumblebee and metric excitations.
However,
$\cE_\mu$ and $\cE^\prime_\mu$ take the same gauge-fixed form 
in the gauge \rf{haxial}.

The two alternative expansions are useful because 
the excitations $\cE_\mu$ and $\cE^\prime_\mu$ are invariant 
under infinitesimal local particle Lorentz transformations.
In these variables,
the local Lorentz symmetry is hidden 
and only the diffeomorphism symmetry is manifest.
The Lorentz NG modes lie in the $b_\mu$-tranverse components
of $\cE_\mu$ or $\cE^\prime_\mu$.
The massive mode $\be$ joins the diffeomorphism mode
in the longitudinal component of $\cE_\mu$.
It can be identified from Eq.\ \rf{redef1} as the combination
\beq
\be =  \fr
{b^\mu (\cE_\mu - \half h_{\mu\nu} b^\nu)}
{b^\al\hat b_\al}
= \mp \hat b^\mu (\cE_\mu - \half h_{\mu\nu} b^\nu) . 
\label{massep}
\eeq
Notice that $\be$ appears here as a
diffeomorphism-invariant combination of field components
that transform nontrivially.

The choice of diffeomorphism gauge has interesting
consequences for $\cE_\mu$ and $\cE^\prime_\mu$.
First,
note that $\cE_\mu$ cannot be gauged to zero,
since only one degree of freedom $\xi_\al b^\al$ appears in
its transformation law.
However,
it is possible to gauge $\cE^\prime_\mu$ to zero.
The corresponding gauge condition for $\cE_\mu$ 
is $\cE_\mu = h_{\mu\nu} b^\nu$.
In either of these gauges,
the massive mode becomes
\beq
\be = \mp \half \hat b^\mu h_{\mu\nu} \hat b^\nu ,
\label{massh}
\eeq
thereby becoming part of the metric.
The bumblebee field strength reduces to
\beq
B_{\mu\nu} = 
(\prt_\mu h_{\nu\si} - \prt_\nu h_{\mu\si}) b^\si ,
\label{Bmunuh}
\eeq
and so is also given by the metric.
Evidently,
in these gauges the theory is defined entirely
in terms of the metric excitations $h_{\mu\nu}$.
Requiring $\cE_\mu$ to vanish 
or dropping the term
$h_{\mu\nu} b^\nu$ in Eq.\ \rf{epprimedown} 
therefore improperly sets to zero
the bumblebee field strength $B_{\mu\nu}$,
which alters the equations of motion governing
the Lorentz NG and massive modes
\cite{bmg}.
It follows from Eq.\ \rf{Bmunuh}
that the effective Lagrange density 
can contain additional kinetic terms for $h_{\mu\nu}$,
beyond those arising from the gravitational action,
that originate from the bumblebee kinetic terms $\cl_{\rm K}$.
These additional terms propagate the Lorentz NG modes,
disguised as massless modes contained in the metric.

\section{Examples}
\label{Examples}

In this section,
some simple examples are developed
to illustrate and enrich the general results obtained 
in the discussions above.
We choose to work within the class of \ks\ bumblebee models
with Lagrange density given by 
Eq.\ \rf{ksbb},
which avoids 
\it a priori \rm 
propagating ghost fields
and the complications of nonminimal gravitational couplings. 
For definiteness,
we set $\La=0$ and 
choose the matter-bumblebee coupling to be
\beq
\cl_{\rm M} = 
- e B_\mu J_{\rm M}^\mu ,
\eeq
where $J_{\rm M}^\mu$ is understood to be a current
formed from matter fields in the theory.
In principle,
this current could be formed from dynamical fields
in the theory or could be prescribed externally.
For the latter case,
diffeomorphism invariance is explicitly broken
unless the current satisfies a suitable differential constraint.

The section begins by
presenting some general results for this class of models,
including ones related to the equations of motion,
conservation laws,
and the connection to Einstein-Maxwell electrodynamics. 
We then turn to a more detailed discussion
of three specific cases
with different explicit potentials $V(X)$,
where the bumblebee field combination $X$ 
is defined as in Eq.\ \rf{XB}.
The three cases involve 
the smooth quadratic potential 
$V_S(X)$ of Eq.\ \rf{VS},
the linear Lagrange-multiplier potential
$V_L(\la, X)$ of Eq.\ \rf{VL},
and the quadratic Lagrange-multiplier potential
$V_Q(\la, X)$ of Eq.\ \rf{VQ}.
The massive modes are studied for each potential,
and their effects on gravity and electromagnetism are explored.

\subsection{General considerations}
\label{General considerations}

\subsubsection{Equations of motion and conservation laws}
\label{Equations of motion and conservation laws}

Varying the Lagrange density \rf{ksbb}
with respect to the metric
yields the gravitational equations of motion 
\bea
G^{\mu\nu} &=& 8 \pi G T^{\mu\nu} .
\label {geq}
\eea
In this expression,
the total energy-momentum tensor
$T^{\mu\nu}$ can be written as a sum of two terms,
\beq
T^{\mu\nu} = T^{\mu\nu}_{\rm M }+ T^{\mu\nu}_B .
\label{Ttotal}
\eeq
The first is the energy-momentum tensor 
$T^{\mu\nu}_{\rm M}$ 
arising from the matter sector.
The second is the energy-momentum tensor 
$T^{\mu\nu}_B$ 
arising from the bumblebee kinetic and potential terms,
\beq
T^{\mu\nu}_B = 
T^{\mu\nu}_{\rm K}
+ T_V^{\mu\nu} ,
\label{TBmunu}
\eeq
where $T^{\mu\nu}_{\rm K}$ and $T_V^{\mu\nu}$ 
are given by 
\bea
T^{\mu\nu}_{\rm K} &=& 
B^{\mu\al} B^\nu_{\pt{\nu}\al} 
- \frac 1 4 g^{\mu\nu} B_{\al\be} B^{\al\be}
\label{tk}
\eea
and 
\bea
T_V^{\mu\nu} &=&
- V g^{\mu\nu} + 2 V^\prime B^\mu B^\nu .
\label{tv}
\eea

Varying instead with respect to the bumblebee field
generates the remaining equations of motion,
\bea
D_\nu B^{\mu\nu} &=& J^\mu.
\label{Beq}
\eea
Like the total energy-momentum tensor,
the total current $J^\mu$ can be written as the sum of two terms
\beq
J^\mu = J_{\rm M}^\mu + J_B^\mu .
\eeq
The partial current $J_{\rm M}^\mu$ 
is associated with the matter sector
and acts as an external source for the bumblebee field.
The partial current $J_B^\mu$ 
arises from the bumblebee self-interaction,
and it is given explicitly as
\beq
J_B^\mu = - 2 V^\prime B^\mu .
\label{JB}
\eeq

The contracted Bianchi identities for $G_{\mu\nu}$
lead to conservation of the total energy-momentum tensor,
\beq
D_\mu T^{\mu\nu} \equiv 
D_\mu (T^{\mu\nu}_{\rm M} + T^{\mu\nu}_B) = 0 .
\label{Tconsv}
\eeq
The antisymmetry of the bumblebee field strength $B_{\mu\nu}$
implies the remnant constraint \rf{remconst}
and hence a second conservation law,
\beq
D_\mu J^\mu \equiv
 D_\mu (J_{\rm M}^\mu + J_B^\mu) = 0 .
\label{Jconsv}
\eeq
Note that this second conservation law is a 
special feature of \ks\ bumblebee models.
It is a direct consequence of choosing 
the \it a priori \rm ghost-free action \rf{ksbb}.
Note also that this conservation law holds 
even though the potential term $V$
excludes any local U(1) gauge symmetry 
in these models.

\subsubsection{Bumblebee currents}
\label{Bumblebee currents}

The bumblebee current $J_B^\mu$ defined in Eq.\ \rf{JB}
vanishes when $V^\prime = 0$.
This situation holds for the vacuum solution and for NG modes,
and it then follows from Eq.\ \rf{Jconsv} 
that the matter current $J_{\rm M}^\mu$ 
is covariantly conserved.
However,
$V^\prime \ne 0$ in the presence of a nonzero massive mode 
or a nonzero linear Lagrange multiplier,
whereupon the bumblebee current $J_B^\mu$
acts as an additional source 
for the bumblebee field equation \rf{Beq}.
A nonzero massive mode or nonzero linear Lagrange multiplier
contributes to the bumblebee component $T^{\mu\nu}_B$ 
of the energy-momentum tensor too,
and it therefore also acts as an additional source
for the gravitational field equations \rf{geq}.

The contributions to the energy-momentum tensor $T^{\mu\nu}_B$ 
stemming from $V^\prime \ne 0$ can be positive or negative.
This is a generic feature,
as can be seen from the expression
for $T_V^{\mu\nu}$ in Eq.\ \rf{TV}
and the general bumblebee Lagrange density \rf{bb}.
The full energy-momentum tensor $T^{\mu\nu}$ is conserved,
but the prospect of negative values for $T^{\mu\nu}_B$ 
implies stability issues for the models.
Under suitable circumstances,
for example,
unbounded negative values of $T^{\mu\nu}_B$ 
can act as unlimited sources of energy 
for the matter sector.
Stability is also a potential issue 
in the absence of matter.
For example,
for the case where $J_{\rm M}^\mu$ is disregarded
and the potential involves a linear Lagrange multiplier 
producing $\la \ne 0$ on shell,
at least one set of initial values is known 
that yields a negative-energy solution
\cite{cj}.

Whether instabilities occur in practice depends on 
the form of the Lagrange density and on the initial conditions.
The situation is comparatively favorable for \ks\ bumblebee models
because the extra conservation law \rf{Jconsv}
can play a role.
In principle,
negative-energy contributions in the initial state 
can be eliminated by choosing initial conditions
such that $V^\prime = 0$,
while the conservation law can prevent the
eventual development of a destabilizing mode.
One way to see this is to expand 
the conservation law Eq.\ \rf{Jconsv}
at leading order in a Minkowski background.
For nonzero $\vev{B_0}$,
we obtain 
\beq
\prt_0 V^\prime \approx
\fr 1 {2 B^0} ( \prt_\mu J_{\rm M}^\mu 
- 2 V^\prime \prt_\mu B^\mu - 2 B^j \prt_j V^\prime ) .
\label{d0Vprime}
\eeq
Suppose the matter current is independently conserved, 
$D_\mu J_{\rm M}^\mu \approx \prt_\mu J_{\rm M}^\mu = 0$.
The initial condition $V^\prime = 0$ 
then implies $\prt_0 V^\prime = 0$ initially. 
Taking further derivatives shows that all time derivatives 
of $V^\prime$ vanish initially and
so $V^\prime$ remains zero for all time,
indicating stability is maintained.

It is interesting to note that
the matter-current conservation law $D^\mu J^{\rm M}_\mu = 0$
emerges naturally in the limit 
of vanishing NG and massive modes,
where the bumblebee field reduces to
its vacuum value $B_\mu \rightarrow \vev{B_\mu} = b_\mu$.
This agrees with a general conjecture
made in Ref.\ \cite{akgrav}.
The line of reasoning is as follows.
In the limit,
the bumblebee theory with spontaneous Lorentz violation
takes the form of a theory with explicit Lorentz violation
with couplings only to $b_\mu$.
However,
theories with explicit Lorentz violation
contain an incompatibility between the Bianchi identities
and the energy-momentum conservation law.
The incompatibility is avoided for spontaneous Lorentz violation 
by the vanishing of a particular variation in the action,
which reduces in the limit to the requirement of current 
conservation $D^\mu J^{\rm M}_\mu = 0$.
Nonetheless,
if $B_\mu$ is excited away from this limit, 
current conservation in the matter sector may fail to hold.

In practice,
the potential instability may be irrelevant for physics.
For example,
if a bumblebee model is viewed 
as an effective field theory emerging 
from an underlying stable quantum theory of gravity,
any apparent instabilities may reflect incomplete
knowledge of constraints on the massive modes
or of countering effects that come into play 
at energy scales above those of the effective theory.
Under these circumstances,
in \ks\ bumblebee models
it may suffice for practical purposes simply to postulate 
that the matter and bumblebee currents do not mix 
and hence obey separate conservation laws,
\beq
D_\mu J_{\rm M}^\mu = 0 ,
\qquad
D_\mu J_B^\mu = 0 .
\label{JJB0}
\eeq
For instance,
one can impose that only matter terms $\cL_M$ 
with a global U(1) symmetry are allowed,
as is done for the case of Minkowski spacetime
in Ref.\ \cite{cfn}. 
Another option might be to disregard $J^{\rm M}_\mu$ altogether
and allow couplings between the matter
and bumblebee fields only through gravity.
In what follows,
we adopt Eqs.\ \rf{JJB0} as needed 
to investigate modifications of gravity 
arising from massive modes
and to study the conditions 
under which Einstein-Maxwell solutions 
emerge in bumblebee models.

\subsubsection{Bumblebee electrodynamics}
\label{Bumblebee electrodynamics}

An interesting aspect of the \ks\ bumblebee models
is their prospective interpretation as alternative theories 
of electrodynamics with physical signatures of Lorentz violation
\cite{bk}.
In this approach,
basic electrodynamics properties 
such as the masslessness of photons are viewed as consequences 
of spontaneous local Lorentz and diffeomorphism breaking
rather than of exact local U(1) symmetry.
A key point is that \ks\ bumblebee models have no dependence
on second time derivatives of $B_0$.
They therefore feature an additional primary constraint 
relative to the bumblebee models \rf{bb}
with more general kinetic terms.
The primary constraint generates a secondary constraint
in the form of a modified Gauss law,
which permits a physical interpretation of the theory
in parallel to electrodynamics.
This modified Gauss law is unavailable to
other bumblebee models.

In a suitable limit,
the equations of motion \rf{geq} and \rf{Beq}
for the \ks\ bumblebee models
reduce to the Einstein-Maxwell equations
in a fixed gauge for the metric and photon fields.
To demonstrate this,
we work in asymptotically Minkowski spacetime
and choose the gauge in which the
bumblebee field is given by Eq.\ \rf{bba}.
The limit of interest is $\be \to 0$,
corresponding to zero massive mode.
For the case with the Lagrange-multiplier potential \rf{VL},
where $\be$ is constrained to zero,
an equivalent result also follows 
in the limit $\la \to 0$.
Since $T_V^{\mu\nu}$ acquires nonzero contributions
only from the massive mode or linear Lagrange multiplier,
it follows that as $\be \to 0$ or $\la \to 0$
the bumblebee energy-momentum tensor $T^{\mu\nu}_B$ 
reduces to that of Einstein-Maxwell electrodynamics,
\bea
T^{\mu\nu}_B &\to& 
T^{\mu\nu}_{\rm K}  
\equiv T^{\mu\nu}_{\rm EM}  
= F^{\mu\al} F_{\al}^{\pt{\al}\nu}
- \frac 1 4 g_{\mu\nu} F_{\al\be} F^{\al\be} .
\qquad 
\label{TBBEM}
\eea
In this equation,
the bumblebee field strength is 
reinterpreted as the electromagnetic field strength
\beq
B_{\mu\nu} \equiv F_{\mu\nu} = 
\prt_\mu A_\nu - \prt_\nu A_\mu ,
\eeq
via the mode expansion \rf{bba},
and the gravitational field equations \rf{geq}
reduce to their Einstein-Maxwell counterparts.
Similarly,
the bumblebee current $J_B^\mu$ in Eq.\ \rf{JB}
vanishes as $\be \to 0$ or $\la \to 0$
because $V^\prime \to 0$.
This leaves only the conventional matter current $J_{\rm M}^\mu$,
which by virtue of Eq.\ \rf{Jconsv} obeys covariant conservation.
The bumblebee field equations \rf{Beq}
therefore also reduce to their Einstein-Maxwell counterparts.
 
Even when $\be$ vanishes,
the interpretation of the model as bumblebee electrodynamics
can in principle be distinguished
from conventional Einstein-Maxwell electrodynamics
through nontrivial SME-type couplings 
involving the vacuum value $b_\mu$
in the matter equations of motion.
Depending on the types of couplings appearing in 
the matter-sector Lagrange density $\cL_{\rm M}$,
a variety of effects could arise.
At the level of the minimal SME,
for example,
axial vector couplings involving $b_\mu$ 
can be sought in numerous experiments,
including ones with 
electrons \cite{eexpt,eexpt2},
protons and neutrons \cite{ccexpt,spaceexpt},
and muons \cite{muexpt}.
Other types of SME coefficients can also be generated
from the vacuum value $b_\mu$.
For example, 
the nonzero symmetric traceless SME coefficients $c_{\mu\nu}$
might emerge via the symmetric traceless product
$b_\mu b_\nu - b^\al b_\al \et_{\mu\nu}/4$,
and experimental searches for these include ones with  
electrons \cite{eexpt,eexpt3},
protons, and neutrons \cite{ccexpt,spaceexpt}.

If the model conditions permit the field $\be$ to be nonzero,
further deviations from Einstein-Maxwell electrodynamics can arise.
For example,
in the weak-field regime
where $\be\hat b_\mu$ is small compared to $b_\mu$,
the bumblebee current $J_B^\mu$ is linear in $\be$ 
and acts as an additional source in the equation
of motion \rf{Beq} for the field strength
$B_{\mu\nu}$.
Effects of this type are investigated
in the next few subsections.
For large fields,
nonlinear effects in the 
bumblebee current and energy-momentum tensor
may also play a role.
However,
the field value of $\be$ is typically suppressed
when the mass of $\be$ is large,
so if the mass is set by the Planck scale 
then the linear approximation is likely to suffice
for most realistic applications.
  
In the limit of zero $\be$,
the physical fields are the field strengths $B_{\mu\nu}$ and
the gravitational field.
Excitations of the bumblebee field $B_\mu$ 
in the classical theory are then unobservable,
in parallel with the potential $A_\mu$ in classical electromagnetism.
This may have consequences for leading-order effects
in the weak-field limit. 
For example,
the weak static limit 
in general bumblebee models \rf{bb}
requires $B_\mu$ itself to be time independent.
However,
at leading order in the \ks\ bumblebee model,
it can suffice to require that only the field strength $B_{\mu\nu}$ 
be time independent
while $B_\mu$ itself has time dependence,
in analogy with Maxwell electrodynamics in Minkowski spacetime.
For example,
the Coulomb electric field $\vec E = - \prt_0 \vec A$
for a static point charge $q$
emerges in temporal gauge $A_0 = 0$ 
from a time-dependent $A_\mu = (0, t \prt_j \Ph_q)$,
where $\Ph_q$ is the Coulomb potential.
Likewise,
in certain leading-order weak static solutions 
for which $\be$ acts as a source of charge, 
the bumblebee field $B_\mu$ may naturally 
exhibit potential lines converging at the source of charge,
similar to the physical configurations
with singular derivatives of $A_\mu$ 
that occur in classical electrodynamics.

Another interesting issue is the role of quantum effects.
Since bumblebee electrodynamics involves gravity,
it faces the same quantization challenges 
as other theories of gravity and electrodynamics,
including Einstein-Maxwell theory.
Some work on the renormalization structure at one loop 
has been performed
\cite{baak,renorm},
and in the limit of zero massive mode and in Minkowski spacetime
the usual properties of quantum electrodynamics
are expected to hold.
Addressing this issue in detail
is an interesting open problem
that lies beyond the scope of this work.

\subsection{Smooth quadratic potential}
\label{Smooth quadratic potential}

This subsection considers the specific \ks\ bumblebee model \rf{ksbb}
having smooth quadratic potential \rf{VS}
with the definition \rf{XB}:
\beq
V = V_S(X) = \half \ka (B_\mu g^{\mu\nu} B_{\nu} \pm b^2)^2.
\label{smoothlagv}
\eeq
For definiteness,
we adopt the mode expansion 
$B_\mu = b_\mu + \cE_\mu$ of Eq.\ \rf{epdown}
in a Minkowski background
and assume weak fields $h_{\mu\nu}$ and $\cE_\mu$
so that the gravitational and bumblebee equations of motion
\rf{geq} and \rf{Beq} can be linearized.

One goal is to investigate deviations from 
Einstein-Maxwell theory arising 
from the presence of a weak nonzero massive mode $\be$. 
We therefore focus on dominant corrections 
to the linearized Einstein-Maxwell equations
arising from $\be$. 
In this approximation,
the bumblebee component of the energy-momentum tensor becomes
\beq
T_B^{\mu\nu} \approx 
T_{\rm EM}^{\mu\nu} + 4 \ka (b^\al \hat b_\al) b^\mu b^\nu \be ,
\label{linTV}
\eeq
where $T_{\rm EM}^{\mu\nu}$ is
the zero-$\be$ limit of $T_{\rm K}^{\mu\nu}$
given in Eq.\ \rf{TBBEM}
and the other term arises from
$T_V^{\mu\nu}$. 
The bumblebee current reduces to
\beq
J_B^{\mu} \approx - 4 \ka (b^\al \hat b_\al) b^\mu \be .
\label{Jbe}
\eeq
The constraint obtained by assuming
current conservation in the matter sector
becomes
\beq
b^\mu \prt_\mu \be \approx 0.
\label{linbeconstraint}
\eeq
These expressions reveal that at leading order
the massive mode $\be$ 
acts as a source for gravitation and electrodynamics,
subject to the constraint \rf{linbeconstraint}.

Note that the linearization procedure 
can alter the time behavior and dynamics of the fields
in the presence of a nonzero massive mode $\be$.
Suppose,
for example,
that $b_\mu$ is timelike
and we adopt the global observer frame in which
$b_\mu = (b,0,0,0)$.
Nonlinearities in the current $J_B^\mu$ 
associated with nonzero $\be$
generate time dependence for most solutions 
because the spatial current $\vec J_B$ 
does work on the field strength $B_{\mu\nu}$.
However,
the linearization \rf{Jbe}
implies $J_B^\mu$ is nonzero only along the direction of $b^\mu$,
reducing to a pure charge density
$J_B^\mu \approx (\rh_B,0,0,0)$
with $\rh_B \approx - 4 \ka b^2 \be$,
and the current-conservation law \rf{linbeconstraint}
then requires $\rh_B$ and hence $\be$ to be static.
Nonetheless,
the linearization procedure captures 
the dominant effects of nonzero $\be$.

\subsubsection{Propagating modes}
\label{Propagating modes}

To study free propagation of the gravitational and bumblebee fields
in the absence of charge and matter,
set $J_{\rm M}^\mu = T_{\rm M}^{\mu\nu} = 0$
in the linearized equations of motion.
The gravitational equations \rf{geq} then become
\bea
&
\square \ol h_{\mu\nu} 
+ \et_{\mu\nu} \prt^\al\prt^\be \ol h_{\al\be}
- \prt^\al\prt_\mu \ol h_{\al\nu} 
- \prt^\al\prt_\nu \ol h_{\al\mu}
\nonumber\\
&
\hskip 20pt
\approx
- 64 \pi G \ka (b^\al \hat b_\al) b_\mu b_\nu \be .
\label{grav3}
\eea
Contributions from $T_{\rm EM}^{\mu\nu}$
are second-order in $\cE_\mu$ and can be neglected 
in this context.
The bumblebee equations \rf{Beq} reduce to 
\beq
\square \cE_\mu - \prt_\mu \prt^\nu \cE_\nu
\approx 4 \ka (b^\al \hat b_\al)  b_\mu \be .
\label{epeq2}
\eeq
Note that the massive mode $\be$
is defined by Eq.\ \rf{massep}
in terms of $\cE_\mu$ and $h_{\mu\nu}$,
so it is not an independent field  
in these equations.
Note also that 
the linearized current-conservation law
\rf{linbeconstraint}
follows by taking a derivative of Eq.\ \rf{epeq2}.

To investigate the behavior of the massive mode $\be$,
it is convenient to combine the above two equations 
to obtain
\beq
\square \be
- 4 \ka (b^\al b_\al) (1 + 4 \pi G b^\be b_\be ) \be 
\approx
\fr 1 {b^\ga \hat b_\ga} b^\la \prt_\la \prt^\mu 
(\cE_\mu - \ol h_{\mu\nu} b^\nu).
\label{beeq}
\eeq
At first glance,
this equation might suggest
that $\be$ can be a propagating massive field. 
However,
$\be$ depends on the fields appearing
on the right-hand side,
so the modes are still coupled and
care is required in determining the dispersion properties,
including the mass value.

A suitable choice of diffeomorphism gauge clarifies matters.
The modes remain entangled in the harmonic gauge \rf{harm}
and also in the barred axial gravitational gauge 
$\ol h_{\mu\nu} b^\nu = 0$.
However,
a sufficient decoupling can be achieved 
by adopting the axial gravitational gauge \rf{haxial}
and by decomposing $\cE_\mu$ into pieces 
parallel and perpendicular to $b_\mu$.
In this gauge,
the longitudinal component $\cE$ of $\cE_\mu$ 
reduces to $\cE = \be$,
so that
\bea
\cE_\mu &=& \cE^t_\mu + \cE \hat b_\mu 
= A_\mu + \be \hat b_\mu .
\eea
In this equation,
we write $\cE^t_\mu \equiv A_\mu $ 
for the transverse components of $\cE_\mu$ 
to make easier the task of tracking
$\be$-dependent deviations 
from conventional electrodynamics.
These components satisfy the axial condition
$A_\mu b^\mu= 0$.
We emphasize that the axial condition 
is \it not \rm a gauge choice.
It is a consequence of projecting along $b_\mu$
and is independent of the gravitational gauge fixing.
It does, however, have the same mathematical form 
as an axial-gauge condition in electrodynamics, 
even though the bumblebee models have no U(1) symmetry.

With these choices,
the usual Einstein equations for 
$\ol h_{\mu\nu}$ 
in axial gravitational gauge
are recovered from the result \rf{grav3} 
when $\be$ vanishes
\cite{bk}.
For nonzero $\be$,
only one linearly independent combination
of the Einstein equations is changed.
It can be written
\bea
\square \ol h 
+ 2 \prt^\mu\prt^\nu \ol h_{\mu\nu}
&\approx&
- 64 \pi G \ka (b^\al \hat b_\al) (b^\be b_\be) \be .
\label{grav4}
\eea
Since the usual propagation-transverse components of $\ol h_{\mu\nu}$ 
remain unaffected,
the physical graviton modes propagate normally.
The bumblebee equations \rf{epeq2} become 
\beq
\square A_\mu - \prt_\mu \prt^\nu A_\nu
+ \fr 1 {b^\be  b_\be} {b_\mu b^\al } \prt_\al \prt^\nu A_\nu
\approx 0 
\eeq
and
\beq
\square \be
- 4 \ka (b^\al b_\al) \be 
\approx
\fr 1 {b^\be \hat b_\be} b^\al \prt_\al \prt^\mu A_\mu .
\label{axbeq}
\eeq
When the massive mode $\be$ vanishes,
these equations reduce to those of electrodynamics
in axial gauge
\cite{bk}.
A nonzero value of $\be$ modifies the equations,
but the usual propagation-transverse components of $A_\mu$
and hence the physical photon modes remain unaffected.

Since $\be$ and $A_\mu$ are independent fields,
the dispersion properties of the massive mode $\be$ 
can be identified from Eq.\ \rf{axbeq}.
Any solutions of the theory 
describing freely propagating modes 
are required to obey the boundary conditions 
that the spacetime be asymptotically flat 
and that the bumblebee fields vanish at infinity.
At linear order,
these solutions are formed from harmonic plane waves
with energy-momentum vectors $p^\mu = (E,\vec p)$ 
obeying suitable dispersion relations.
The modes satisfying the asymptotic boundary conditions 
are then constructed as Fourier superpositions
of the plane-wave solutions.
For the massive mode $\be$,
the constraint equation \rf{linbeconstraint}
imposes the additional requirement
\beq
b^\mu p_\mu \approx 0 .
\label{bk}
\eeq
Any freely propagating massive mode 
is therefore constrained to have an energy-momentum vector 
orthogonal to the vacuum value $b_\mu$.
For harmonic-mode solutions to the equations of motion,
this requirement implies that the term on the right-hand side 
of Eq.\ \rf{axbeq} vanishes in Fourier space.
The resulting dispersion law for the massive mode $\be$ 
is then 
\beq
p^\mu p_\mu
\approx - 4 \ka b^\al b_\al ,
\label{kdisp}
\eeq
which involves the squared-mass parameter 
\beq
M_\be^2 = 4 \ka b^\al b_\al .
\label{bemass}
\eeq
The sign of $M_\be^2$ 
depends on whether $b_\mu$ is timelike or spacelike.
The scale of $M_\be^2$ depends on both $\ka$ and $b$.
Inspection of Eqs.\ \rf{grav4} or \rf{axbeq}
shows that the limits $|M_\be|^2 \to \infty$
or $\ka\to \infty$ with $b_\mu$ fixed 
are equivalent to taking the limit of vanishing
massive mode, $\be\to 0$. 
The discussion in Sec.\ \ref{Bumblebee electrodynamics}
then implies that the limit of large $|M_\be|^2$
approximates Einstein-Maxwell theory.

For the case of a timelike vacuum value $b_\mu$,
the parameter $M_\be^2 = -4\ka b^2$ 
has the wrong sign for a physical mass.
Adopting the global observer frame in which
$b_\mu = (b,0,0,0)$,
we see that the constraint \rf{bk}
forces the energy $E$ of this mode to vanish,
leaving the condition that the magnitude of the momentum 
must remain fixed at 
$|\vec p| = |M_\be|$.
A time-independent mode with fixed spatial wavelength 
cannot be Fourier superposed to form a physical wave packet 
satisfying the asymptotic boundary conditions.
It follows that no physical propagating massive mode appears
when $b_\mu$ is timelike.

If instead the vacuum value $b_\mu$ is spacelike,
then $M_\be^2 = 4\ka b^2$ is positive 
and is a candidate for a physical mass.
Choose the global observer coordinate system
such that the spacelike $b_\mu$ 
takes the form $b_\mu = (0,0,0,b)$.
The constraint \rf{bk} then requires the vanishing 
of the momentum along the spatial direction $b_3$,
and the dispersion law \rf{kdisp} becomes 
\beq
E^2 - (p_x^2 + p_y^2) \approx M_\be^2 .
\label{spacelikedisp}
\eeq
This suggests that $\be$ could propagate 
as a harmonic plane wave
along spatial directions transverse to $b_\mu$.
However,
these harmonic plane waves are constant in $z$
and hence fail to satisfy the 
asymptotic boundary conditions as $z\to \pm \infty$
unless their amplitude is zero.
Since all harmonics propagate under the same constraint \rf{bk},
no Fourier superposition can be constructed
to obey the boundary conditions.
It follows that 
no physical propagating massive mode exists
for the case of a spacelike vector either.

The above results can also be confirmed
by obtaining the eigenvalues
of the Fourier-transformed equations of motion,
which form a 14$\times$14 matrix
determining the nature of the modes.
Consider,
for example,
the case of purely timelike $b_\mu$.
In the axial gravitational gauge,
four of the modes are zero,
corresponding to the four fixed degrees of freedom. 
Another four eigenvalues describe modes propagating 
via a massless dispersion law,
corresponding to the two photon and two graviton modes.
One eigenvalue has zero energy and sets
$|\vec p| = |M_\be|$,
corresponding to the massive mode $\be$.
The remaining 5 modes are auxiliary,
and all have zero energy. 
Related results can be obtained 
for the case of spacelike $b_\mu$.

\subsubsection{Weak static limit}
\label{Weak static limit}

Although the massive mode $\be$ is nonpropagating, 
it plays the role of an auxiliary field
and can thereby produce various effects.
For example,
its presence can affect the weak static limits 
of the gravitational and bumblebee interactions.
In particular,
a nonzero $\be$ can modify the forms of both 
the Newton and Coulomb potentials.

To demonstrate this,
we consider the weak static limit of the field equations
with an external matter sector containing massive point charges.
The matter Lagrange density can be taken as 
\bea
\cL_{\rm M} &=& 
\sum_n m_n \int d\ta
\left[ g_{\mu\nu} 
\fr {dx_n^\mu}{d\ta} \fr {dx_n^\nu}{d\ta} 
\right]^{1/2} 
\de^4 (x - x_n(\ta)) 
\nonumber
\\
&&
+ \sum_n q_n \int d\ta ~ B_\mu(x) 
\fr {dx_n^\mu} {d\ta} \de^4 (x - x_n(\ta)),
\label{LM}
\eea
where $m_n$ and $q_n$ are the masses and charges 
of the point particles,
respectively.
The energy-momentum tensor for the matter sector is
\beq
T_{\rm M}^{\mu\nu} = \fr 1 e \sum_n m_n
\int d\ta \fr {dx_n^\mu} {d \ta}  \fr {dx_n^\nu} {d \ta}
\de^4 (x - x_n(\ta)) .
\label{TMm}
\eeq
Its combination with $T_B^{\mu\nu}$ given in Eq.\ \rf{TBmunu} 
forms the total energy-momentum tensor, which is conserved.
The matter-sector current 
$J_{\rm M}^\mu \equiv (\rh_q, \vec J)$ is
\beq
J_{\rm M}^\mu
= \fr 1 e \sum_n q_n \int d\ta  
\fr {d x_n^\mu} {d \ta} \de^4 (x - x_n(\ta)) .
\label{Jq}
\eeq
It is conserved, $D_\mu J_{\rm M}^\mu = 0$.

The standard route to adopting the weak static limit 
is to linearize the equations of motion
in the fields 
and then discard all time derivatives.
This meets no difficulty under suitable circumstances,
such as the vacuum value $b_\mu$ being purely spacelike,
and we follow this route in the present subsection.
However,
in certain cases additional care is required  
because the $b_\mu$-transverse components of the bumblebee field 
obey the axial condition $b^\mu \cE^t_\mu = 0$,
which can imply time dependence even for static fields.
For example,
as discussed in Sec.\ \ref{Bumblebee electrodynamics},
the case of purely timelike $b_\mu$ and zero $\be$
is equivalent to electrodynamics in temporal gauge,
for which the static Coulomb potential 
involves a linear time dependence in $\vec A$.
The corresponding weak static limit with nonzero $\be$ 
can therefore be expected to exhibit this feature.
We return to this issue in the next subsection. 

To proceed,
it is useful to choose a diffeomorphism gauge
that simplifies the Einstein tensor 
and allows direct comparison 
between the geodesic equation and the Newton force law.
One possible choice is the harmonic gauge \rf{harm},
in which the static gravitational potential $\Ph_g$
is related to the metric by
$\Ph_g = -\half h_{00} = -\half \ol h_{00} + \frac 1 4 \ol h$.
The linearized gravitational equations become
\beq
-\half \square \ol h^{\mu\nu}
\approx
8\pi G
[T_{\rm M}^{\mu\nu}
+ T_{\rm EM}^{\mu\nu} 
+ 4 \ka (b^\al \hat b_\al) b^\mu b^\nu \be] ,
\label{lsharm}
\eeq
where $\be$ is the massive mode 
given in terms of $\cE_\mu$ and $h_{\mu\nu}$
in Eq.\ \rf{massep},
and $T_{\rm EM}^{\mu\nu}$ 
is the zero-$\be$ form of $T_{\rm K}^{\mu\nu}$
given in Eq.\ \rf{TBBEM}.

We can now extract the relevant equations
in the weak static limit.
The equation for the gravitational potential $\Ph_g$
is the 00 component of the trace-reversed form
of Eq.\ \rf{lsharm},
\beq
\nabla^2 \Ph_g 
\approx
8\pi G
[\ol T{}_{\rm M}^{00}
+ \ol T{}_{\rm EM}^{00} 
+ 2 \ka (b^\al \hat b_\al) (b_0^2 + \vec b^{\, 2}) \be] ,
\label{lsharm2}
\eeq
where the trace-reversed energy-momentum tensors
are defined by 
$\ol T_{\mu\nu} \equiv T_{\mu\nu} - \et_{\mu\nu} T^\al_{\pt{\al}\al}$
and are evaluated in the static limit.
The weak static bumblebee equations become 
\bea
\vec \nabla^2 \cE_0 
&\approx & \rh_q
+ 4 \ka (b^\al \hat b_\al)  b_0 \be ,
\nonumber\\
\vec \nabla^2 \vec \cE 
- \vec\nabla (\vec\nabla\cdot\vec \cE) &\approx & \vec J 
+ 4 \ka (b^\al \hat b_\al)\vec b \be ,
\label{stlinkappaeqs}
\eea
where $\rh_q$ and $\vec J$ are static charge and current sources.

The above equations show that
the massive mode $\be$ can be viewed
in the weak static limit
as a simultaneous source of 
energy density, charge density, and current density.
The relative weighting of the contributions
is controlled by the relative sizes of the components of $b_\mu$.
In fact,
the same is also true of contributions
to gravitomagnetic and higher-order gravitational effects
arising from other components of the gravitational equations.
In the interpretation as bumblebee electrodynamics,
the equations show that we can expect
deviations from the usual Newton and Coulomb force laws
when $\be$ is nonzero.
Since the effects are controlled by components of $b_\mu$,
they are perceived differently from different observer frames,
much like charge and current in ordinary electrodynamics.
Also,
the unconventional source terms are linear in $\be$,
so the solutions can be interpreted as superpositions
of conventional gravitational and electrodynamic fields 
with effects due to $\be$.
However, 
since $\be$ itself is formed as the combination \rf{massep}
of the metric and bumblebee fields,
the extra source terms in fact reflect
contributions from the gravitational and electrodynamic fields 
that are absent in the Einstein-Maxwell limit.

\subsubsection{Timelike case}
\label{Timelike case}

In this subsection,
we consider in more detail the weak static limit
with timelike $b_\mu$.
We solve explicitly the weak static equations 
in the special observer frame
for which $b_\mu = (b,0,0,0)$ is purely timelike,
and we provide some remarks about the more general case.
The special frame might be identified 
with the natural frame of the cosmic microwave background.
However,
in laboratory searches for modifications 
to weak static gravitation and electrodynamics,
this frame can at best be an approximation
because the Earth rotates and revolves about the Sun,
which in turn is moving 
with respect to the cosmic microwave background.
Nonetheless,
the explicit solutions obtained here
offer useful insight into the modifications 
to the Newton and Coulomb potentials
that are introduced by the massive mode.

For the purely timelike case,
both $h_{\mu\nu}$ and $B_{\mu\nu}$
are expected to be time independent in the weak static limit. 
However, 
time dependence must appear in $\vec \cE$
even in the static limit
because as $\be \to 0$
the match to electrodynamics
arises in the temporal gauge, 
for which the standard Coulomb solution 
has a linear time dependence.
For this reason,
in taking the weak static limit in what follows,
we keep time derivatives 
acting on the vector components $\vec \cE$.

The analysis for purely timelike $b_\mu$
could be performed in the harmonic gauge,
but it turns out that 
choosing the transverse gauge instead
results in the complete decoupling from the massive mode
of nine of the 10 gravitational equations.
The transverse gauge is defined by
\beq
\prt^j h_{0j} = 0, 
\qquad
\prt^j h_{jk} = \frac 1 3 \prt_k h^j_{\pt{j}j}, 
\label{htransverse}
\eeq
where $j$, $k$ range over the spatial directions.
In this gauge,
$G_{00} \approx - \nabla^2 h_{00} = 2 \nabla^ 2 \Ph_g$,
where $\Ph_g$ is the gravitational potential.

Taking the weak static limit of the gravitational and bumblebee
equations of motion in this gauge yields
\bea
\nabla^2 \Ph_g 
&\approx&
4\pi G
(T_{\rm M}^{00} + T_{\rm EM}^{00} - 4 \ka b^3 \be) ,
\nonumber\\
\vec \nabla^2 \cE_0 - \vec \nabla \cdot \prt_0 \vec \cE
&\approx & \rh_q
- 4 \ka b^2 \be ,
\nonumber\\
\square \vec \cE 
- \vec\nabla (\vec\nabla\cdot\vec \cE) &\approx & \vec J .
\label{trlinkappaeqs}
\eea
The other nine components of $h_{\mu\nu}$
obey equations that are identical to those
of general relativity in the weak static limit. 
In Eq.\ \rf{trlinkappaeqs},
$T_{\rm M}^{00}$ is the static energy density 
in the matter sector,
and $T_{\rm EM}^{00}$ is the static $00$ component 
of the zero-$\be$ limit of $T_{\rm K}^{\mu\nu}$
given in Eq.\ \rf{TBBEM}.
The charge density $\rh_q$ and current density $\vec J$
are also understood to be time-independent distributions.

The explicit form of the massive mode $\be$ 
follows from Eq.\ \rf{massep}
and is found to be
\beq
\be = \cE_0 - b \Ph_g .
\label{betimelike}
\eeq
For purely timelike $b_\mu$,
the constraint \rf{linbeconstraint} reduces to 
\beq
\prt_0 \be \approx 0,
\label{timeindepbe}
\eeq
so $\be$ is time independent at this order.
As described in the previous subsection,
$\be$ can be interpreted as a source 
for $\Ph_g$ and $\cE_0$.
Here,
$\be$ acts as a static source 
of energy density and charge density with 
\bea 
T_B^{\mu\nu} &=& b \rh_B \et^{0\mu} \et^{0\nu} ,
\nonumber\\
J_B^\mu &=& (\rh_B,0,0,0),
\label{betastsource}
\eea
where $\rh_B = - 4 \ka b^2 \be$.

To find the explicit modifications 
to the gravitational and electromagnetic potentials 
caused by $\be$,
we focus on the case of a point particle 
of mass $m$ and charge $q$
located at the origin.
The source terms in this case
have the form
\bea
T_{\rm M}^{00} &=& \rh_m = m \de^{(3)}(\vec x) , 
\nonumber\\
J_{\rm M}^0 &\equiv& \rh_q = q \de^{(3)}(\vec x) , 
\label{pointsource}
\eea
and all other components of 
$T_{\rm M}^{\mu\nu}$ and $J_{\rm M}^\mu$ are zero.
In the absence of Lorentz violation,
the weak static solution to the Einstein-Maxwell equations
with these source terms
consists of the linearized Reissner-Nordstr\"om metric
and corresponding electromagnetic fields.
Denote by $\Ph_m$ the associated gravitational potential.
It obeys
\beq
\vec \nabla^2 \Ph_m = 
4 \pi G (\rh_m + T_{\rm EM}^{00} ).
\label{RN}
\eeq
In the limit $q \rightarrow 0$,
it reduces to the Newton gravitational potential and
obeys the usual Poisson equation
$\nabla^2 \Ph_m = 4 \pi G \rh_m$.
Similarly,  
denote the conventional Coulomb potential by $\Ph_q$.
It obeys the Poisson equation
\beq
\nabla^2 \Ph_q = - \rh_q.
\label{fish}
\eeq

It is also convenient to express 
the massive mode $\be$ in terms of a potential $\Ph_B$.
We define
\beq
\be \equiv 
\fr 1 {| M_\be |^2}
\vec \nabla^2 \Ph_B (\vec x) ,
\label{Phbe}
\eeq
where $| M_\be |^2$ is the absolute value 
of the squared mass in Eq.\ \rf{bemass}.
Note that $\Ph_B$
is time-independent due to the constraint 
\rf{timeindepbe}.
Since $\be$ acts as a source of static charge density 
given by Eq.\ \rf{betastsource},
the definition of $\Ph_B$ can be interpreted 
as a third Poisson equation,
\beq
\vec\nabla^2 \Ph_B = - \rh_B .
\eeq
The time independence of both $\be$ and $\Ph_B$ 
means they are determined by their initial values.

For the point particle of mass $m$ and charge $q$,
the general solution for 
the gravitational potential $\Ph_g$ 
and the bumblebee field $\cE_\mu$
in the weak static limit 
can be expressed in terms of the conventional
potentials $\Ph_m$ and $\Ph_q$
and the bumblebee potential $\Ph_B$.
We obtain  
\bea
\Ph_g &=& 
\Ph_m - 4\pi G b \Ph_B ,
\nonumber\\
\cE_0 &=& 
b \Ph_m - 4\pi G b^2 \Ph_B 
+ \fr 1 {| M_\be |^2} \vec \nabla^2 \Ph_B ,
\nonumber \\
\cE_j &=& t \prt_j [
\Ph_q + b \Ph_m 
+(1 - 4 \pi G b^2 ) \Ph_B  
+ \fr 1 {| M_\be |^2} \vec \nabla^2 \Ph_B 
].
\nonumber \\
\label{Phisol}
\eea
The corresponding 
static gravitational field $\vec g$,
static electric field $\vec E$,
and static magnetic field $\vec B$
are found to be 
\bea
\vec g &=& - \vec \nabla \Ph_m + 4\pi G b \vec \nabla \Ph_B ,
\nonumber\\
\vec E &=& - \vec \nabla \Ph_q - \vec \nabla \Ph_B ,
\nonumber\\
\vec B &=& 0 .
\label{EB}
\eea
The static Maxwell equations include a modified Gauss law
\bea
\vec \nabla \cdot \vec E &=& 
- \vec \nabla^2 \Ph_q - \vec \nabla^2 \Ph_B
\nonumber \\
&=& \rh_q + \rh_B ,
\label{gauss}
\eea
together with the usual static laws 
\bea
\vec \nabla \cdot \vec B &=& 0,
\nonumber\\
\vec \nabla \times \vec E &=& 0,
\nonumber\\
\vec \nabla \times \vec B &=& 0.
\eea
These equations demonstrate for the purely timelike case
the ways in which a nonzero massive mode $\be$ modifies 
the conventional static gravitational and electrodynamic fields. 

Compared to electrodynamics,
the purely timelike \ks\ bumblebee model 
introduces an additional degree of freedom
into the problem of determining 
the static electric field $\vec E$.
This extra degree of freedom is the massive mode $\be$,
and the extent of its effects depends
on the initial conditions for $\be$.
In the absence of 
a satisfactory underlying theory predicting $\be$
or of direct experimental observation of $\be$,
these initial conditions are undetermined.

One natural choice is to adopt $\Ph_B = 0$ 
as an initial condition,
which implies $\Ph_B (\vec x) \approx 0$ for all time
and hence corresponds to zero massive mode $\be$.
The solution \rf{Phisol}
then reduces to the weak static limit of 
Einstein-Maxwell theory,
as expected.
The gravitational potential becomes $\Ph_g = \Ph_m$.
The bumblebee field reduces to 
$\cE_\mu = (b \Ph_m, t \prt_j (\Ph_q + b \Ph_m))$,
which despite the appearance of the gravitational potential
implies $\vec E = - \vec \nabla \Ph_q$ 
and the usual Gauss and Poisson equations 
$\vec \nabla \cdot E = - \vec \nabla^2 \Ph_q = \rh_q$.
We note in passing that conventional electrodynamics
is also recovered in the limit $|M_\be|^2 \rightarrow \infty$
because taking this limit in Eq.\ \rf{trlinkappaeqs} 
implies $\be \to 0$ and hence $\Ph_B \to 0$.
Thus,
even with a nonzero massive mode $\be$,
a theory with a large mass parameter $|M_\be|$
approximates Einstein-Maxwell theory in the low-energy regime,
in agreement with the discussion following Eq.\ \rf{bemass}.

Other choices of initial conditions on $\Ph_B$
might appear in the context of a more fundamental theory
for which a bumblebee model 
provides an effective partial low-energy theory.
One set of examples with nonzero $\Ph_B$
consists of solutions with zero matter current,
$J_{\rm M}^\mu\equiv 0$.
Imposing $J_{\rm M}^\mu\equiv 0$ by hand
is a common approach in the literature.
It implies the bumblebee field has no couplings to matter
and hence is unrelated to electrodynamics,
being a field that interacts only gravitationally.
The massless Lorentz NG modes then propagate freely 
as sterile fields that carry energy and momentum
but convey no direct forces on charged particles.
We emphasize here that a theory of this type
can nonetheless lead to modifications of the 
gravitational interaction,
contrary to some claims in the literature.
This is readily illustrated 
in the context of the purely timelike example
considered in this subsection.
With $J_{\rm M}^\mu\equiv 0$,
the linearized solutions for a static point particle of mass $m$
are still given by Eqs.\ \rf{Phisol},
but $\Ph_q = 0$ and the fields $\vec E$ and $\vec B$ 
are largely irrelevant.
However,
the gravitational potential $\Ph_g$ is modified 
by the potential $\Ph_B$ for the massive mode $\be$,
with the value of $\Ph_B$ fixed by initial conditions.

In the absence of additional
theoretical or experimental information,
the possibilities for model building are vast. 
One could, 
for example, 
consider initial conditions with $\Ph_B$ proportional to $\Ph_m$,
\beq
\Ph_B = \fr \al {4 \pi G b} \Ph_m ,
\label{scalePh}
\eeq
with $\al$ a constant.
For this class of solutions,
$\Ph_g$ has the usual form $\Ph_g = -G^* m/r$
for a point mass 
but with a scaled value of the Newton gravitational constant
\beq
G^* = ( 1 - \al ) G_N ,
\label{Gstar}
\eeq
where $G_N$ is the value of $G^*$ for $\Ph_B = 0$.
This rescaling has no observable effects 
in the special observer frame 
because laboratory measurements 
would determine $G^*$ instead of $G_N$,
although sidereal, annual, and solar motions
might introduce detectable effects in realistic experiments.
One could equally well consider instead examples 
in which $\Ph_B$ has 
nonlinear functional dependence on $\Ph_m$
or has other coordinate dependence.
Various forms could be proposed as candidate models 
to explain phenomena such as dark matter 
or possibly dark energy.
With the choice of $\Ph_B$ being conjecture
at the level of the linearized equations,
this approach is purely phenomenological.
However,
if a bumblebee model appears 
as part of the effective theory 
in a complete theory of quantum gravity,
the form of $\Ph_B$ could well be fixed
and predictions for dark matter and perhaps dark energy
could become possible. 

We conclude this subsection with some comments 
about the case where $b_\mu$ takes a more general timelike
form in the given observer frame $O$.
To investigate this,
one can either study directly the weak static limit 
of the gravitational and bumblebee equations
with a charged massive point particle at rest at the origin
in $O$ as before,
or one can boost the observer frame $O$
to another frame $O^\prime$
in which $b_\mu$ is purely timelike 
but the point particle is moving.
The two pictures are related by observer Lorentz transformations 
and are therefore physically equivalent.

In the frame $O$ in which the particle is at rest,
the matter energy-momentum tensor and current 
are given by Eq.\ \rf{pointsource} as before,
but the bumblebee energy-momentum tensor $T_B^{\mu\nu}$
and current $J_B^\mu$
have additional nonzero components 
due to the nonzero spacelike components of $b_\mu$.
The complete solution therefore involves components of $h_{\mu\nu}$
other than the gravitational potential $\Ph_g$.
However,
we can gain insight into the behavior of the massive mode
by considering the constraint \rf{linbeconstraint}
and performing an observer Lorentz boost 
with velocity $\vec v_B = - \vec b/b_0$
to a frame $O^\prime$ in which $b_\mu$ becomes purely timelike.
Since this transformation must reduce 
the constraint \rf{linbeconstraint} 
to the purely timelike form \rf{timeindepbe},
it follows that in the original frame $O$
the constraint takes the form 
\beq
\prt_0 \be + \vec v_B \cdot \vec \nabla \be = 0,
\eeq
which implies
\beq
\be = \be (\vec x - \vec v_B t).
\eeq
We see that the massive mode $\be$ moves in the frame $O$
with a velocity $\vec v_B$ equal to the relative boost velocity 
linking $b_\mu$ to its purely timelike form.
The gravitational and electrodynamic equations
are therefore modified by time-dependent source terms 
moving with velocity $v_B$ relative to the point particle.
An equivalent result is obtained 
by working directly in the frame $O^\prime$,
in which $b_\mu$ is purely timelike
and $\be$ is therefore static,
but in which all components of
the matter energy-momentum tensor and current are nonzero
and represent a particle moving with velocity $-\vec v_B$
relative to $\be$.
As before,
the form of the massive mode $\be$
is set by initial conditions.
Since the equations are linear in either frame,
the solutions consist of a superposition of 
conventional potentials and the massive-mode contributions
with relative motion.

\subsection{Linear Lagrange-multiplier potential}
\label{Linear Lagrange-multiplier potential}

In this subsection,
we discuss the \ks\ bumblebee model \rf{ksbb}
with the linear Lagrange-multiplier potential \rf{VL}
and the definition \rf{XB}:
\beq
V = V_L(\la, X) = \la (B_\mu g^{\mu\nu} B_{\nu} \pm b^2).
\label{linlagv}
\eeq
Bumblebee models with this potential 
have been widely studied at the linearized level
for about two decades 
\cite{ks}.
Here, 
we compare and contrast this theory 
to the case with the smooth quadratic potential $V_S$.
We show the Lagrange multiplier $\la$ 
produces effects in the $V_L$ model 
that are very similar to those 
of the massive mode $\be$ in the $V_S$ model.
Paralleling the treatment of the $V_S$ model
in Sec.\ \ref{Smooth quadratic potential},
the mode expansion 
$B_\mu = b_\mu + \cE_\mu$ of Eq.\ \rf{epdown}
in a Minkowski background is adopted,
and the fields $h_{\mu\nu}$ and $\cE_\mu$ are assumed weak
so that linearization can be performed.

Variation of the action produces
the gravitational and bumblebee equations of motion
\rf{geq} and \rf{Beq},
together with the Lagrange-multiplier constraint 
\beq
X = B_\mu g^{\mu\nu} B_\nu \pm b^2 = 0  .
\label{lmconst}
\eeq
This condition enforces the vanishing of the massive mode,
as discussed in Sec.\ \ref{Excitations},
which leaves only the Lorentz NG modes 
as possible excitations of the bumblebee field $B_\mu$.
However, 
there is also an additional degree of freedom,
the Lagrange multiplier $\la$ itself,
that appears in the equations of motion.

One might naively expect the $V_L$ model to yield solutions 
identical to those obtained 
in the infinite-mass limit $|M_\be| \rightarrow \infty$ 
of the $V_S$ model.
For example,
one might reason that an infinite mass 
would make energetically impossible
any field excitations away from the minimum of $V_S$,
leading to the constraint \rf{lmconst}.
However, 
the potential $V_L$ is a function of two combinations of fields,
$V_L = V_L(\la, X)$,
whereas $V_S(X)$ involves only one.
The infinite-mass limit indeed suppresses $X$ excitations 
away from $X=0$ in $V_S$,
but it contains no match for $\la$.
There is thus an extra field degree of freedom in $V_L$
relative to the infinite-mass limit of $V_S$.
For example,
$V_S^\prime \rightarrow 0$ in the infinite-mass limit 
because no excitations of $X$ are allowed,
whereas $V_L^\prime = \la$,
which need not vanish.
We therefore conclude that
the correspondence between 
the infinite-mass limit of the $V_S$ model
and the $V_L$ model can occur only when $\la = 0$. 

To gain intuition about the effects
associated with the Lagrange-multiplier field $\la$,
consider its role in the equations of motion.
Since the constraint \rf{lmconst} ensures
no massive excitations can appear,
the bumblebee energy-momentum tensor \rf{TBmunu} reduces to 
\beq
T^{\mu\nu}_B = 
T^{\mu\nu}_{\rm EM}
+ T_V^{\mu\nu} .
\label{lmTBmunu}
\eeq
We find
\bea
T_V^{\mu\nu} &=& 2 \la B^\mu B^\nu \approx 2 b^\mu b^\nu \la,
\eea
where the last form is the leading-order contribution
in the linearized limit.
Similarly,
the bumblebee current \rf{JB}
becomes
\bea
J_B^\mu &=& - 2 \la B^\mu \approx -2 b^\mu \la.
\label{TJlm}
\eea
Conservation of the matter current $J_M^\mu$ 
implies the constraint
\beq
b^\mu \prt_\mu \la \approx 0.
\label{lmconstraint}
\eeq
When $\la \to 0$,
all these equations reduce 
to those of Einstein-Maxwell theory,
in agreement with the discussion of bumblebee electrodynamics 
in Sec.\ \ref{Bumblebee electrodynamics}.
Moreover,
by comparison with 
Eqs.\ \rf{linTV}--\rf{linbeconstraint},
we see that the Lagrange multiplier $\la$ in the $V_L$ model
plays a role very comparable 
to that of the massive mode $\be$ in the $V_S$ model.
In effect,
$\la$ acts as an additional source of energy-momentum density
and current density in the equations of motion.
 
The propagating modes for the \ks\ bumblebee model 
with the linear Lagrange-multiplier potential $V_L$ 
have been investigated elsewhere
\cite{ks,bk}.
The usual two graviton modes 
propagate as free transverse massless modes
independently of $\la$,
as do the usual two photon modes
emerging from the Lorentz NG modes.
Since $\la$ has no kinetic terms,
it is auxiliary and cannot propagate.

The weak static limit of the $V_L$ model 
can also be studied,
following the lines of the discussion for the $V_S$ model
in Sec.\ \ref{Timelike case}.
Consider,
for example,
the case of purely timelike $b_\mu = (b,0,0,0)$.
The constraint \rf{lmconstraint} then implies 
that $\la$ must be time independent at leading order,
\beq
\prt_0 \la \approx 0.
\label{lmfixing}
\eeq
A nonzero $\la$ therefore acts 
as an additional static source of energy density
and charge density,
which can modify the static potentials.
We again adopt the transverse gauge \rf{htransverse},
for which the gravitational potential is
$\Ph_g = -h_{00}/2$.
Linearizing the constraint \rf{lmconst}
for the purely timelike case yields
\beq
\cE_0 - b \Ph_g = 0 .
\label{lambdaconstraint}
\eeq
As expected,
this condition corresponds in the context of the $V_S$ model
to enforcing the vanishing of the massive mode $\be$.
In the context of Einstein-Maxwell theory,
it corresponds as before to a gauge-fixing condition
that reduces to the usual temporal gauge
in the absence of gravity.

For a single point particle of charge $q$ and mass $m$
at rest at the origin,
the energy density $\rh_m$ and charge density $\rh_q$ 
are given by Eq.\ \rf{pointsource}.
The equations of motion in the weak static limit become
\bea
\nabla^2 \Ph_g 
&\approx&
4\pi G
(\rh_m + T_{\rm EM}^{00} + 2 b^2 \la ) ,
\nonumber\\
\vec \nabla^2 \cE_0 - \vec \nabla \cdot \prt_0 \vec \cE
&\approx & \rh_q
+ 2 b \la ,
\nonumber\\
\square \vec \cE 
- \vec\nabla (\vec\nabla\cdot\vec \cE) &\approx & 0 .
\label{linlambdaeqs}
\eea
Comparison of these expressions 
with Eqs.\ \rf{trlinkappaeqs} for the $V_S$ model 
again reveals the correspondence between the roles
of the Lagrange-multiplier field $\la$ and the massive mode $\be$. 

With $\la$ acting as an effective source of charge and energy density,
we can introduce a potential $\Ph_B$,
\beq
\la \equiv - \fr 1 {2 b} \nabla^2 \Ph_B  ,
\label{lambdaPhi}
\eeq
which equivalently can be viewed as the solution
to the Poisson equation
\beq
\nabla^2 \Ph_B = - \rh_B ,
\qquad
\rh_B \equiv 2b\la.
\label{lafish}
\eeq
As before,
let $\Ph_m$ denote the usual gravitational potential 
obeying Eq.\ \rf{RN},
and let $\Ph_q$ denote the usual Coulomb potential 
obeying Eq.\ \rf{fish}.
In terms of these three potentials,
the weak-field static solutions 
to the equations of motion \rf{linlambdaeqs} 
can be written as
\bea
\Ph_g &=& 
\Ph_m - 4\pi G b \Ph_B ,
\nonumber\\
\cE_0 &=& 
b \Ph_m - 4\pi G b^2 \Ph_B ,
\nonumber \\
\cE_j &=& t \prt_j [
\Ph_q + b \Ph_m 
+(1 - 4 \pi G b^2 ) \Ph_B  
].
\label{Philambdasol}
\eea
The associated 
static gravitational field $\vec g$,
static electric field $\vec E$,
and static magnetic field $\vec B$
are all given by the same mathematical expressions
as Eqs.\ \rf{EB},
but the potential $\Ph_B$ is now defined in terms of $\la$ 
according to Eq.\ \rf{lambdaPhi}.
Similarly,
a nonzero Lagrange-multipler field $\la$ yields 
the same modified mathematical form \rf{gauss} of the Gauss law,
but the potential $\Ph_B$ and charge $\rh_B$
are now defined in terms of $\la$.

The bumblebee potential $\Ph_B$
in Eqs.\ \rf{Philambdasol}
is undetermined by the equations of motion \rf{linlambdaeqs} 
and must be specified as an initial value.
This specification also fixes 
the initial value of the Lagrange multiplier $\la$,
and the condition \rf{lmfixing}
then ensures that $\la$ remains unchanged for all time.
The situation here parallels that of the $V_S$ model,
where initial conditions must be imposed that 
subsequently fix the bumblebee potential 
and the massive mode $\be$ for all time.
These results are special cases of a more general fact
often overlooked in the literature:
to be well defined,
all bumblebee models require explicit initial conditions
on the massive modes and Lagrange-multiplier fields.
The subsequent development of the massive modes
and Lagrange multipliers can then be deduced 
from the equations of motion or from derived constraints.
In the absence of specified initial conditions,
physical interpretations and predictions 
cannot be reliable.
Moreover,
in the absence of direct experimental observation
or prediction from an underlying theory,
the choice of initial conditions is largely unrestricted
and can lead to widely differing effects. 

One natural initial condition is $\Ph_B = 0$, $\la = 0$.
The above results then reduce to Einstein-Maxwell theory 
in the weak static limit,
as expected.
This $V_L$ model has only field excitations maintaining 
both $V_L=0$ and $V_L^\prime = 0$.
It corresponds to the infinite-mass limit of $V_S$,
which is itself equivalent to Einstein-Maxwell theory.

Another possibility is to consider models 
without direct coupling to matter,
$J_{\rm M}^\mu\equiv 0$,
which implies $\rh_q \equiv 0$, $\Ph_q \equiv 0$ 
in the above equations. 
Like their $V_S$ counterparts 
discussed in Sec.\ \ref{Timelike case},
these models are purely gravitational
and are unrelated to electrodynamics. 
As always, 
initial conditions on $\Ph_B$ must be specified. 
One simple possibility,
for example,
is to choose $\Ph_B = - b \Ph_g$.
This $V_L$ model is contained 
in the analysis of Ref.\ \cite{cli}.
The spatial components of the bumblebee vanish, $\cE_j = 0$,
so the bumblebee field $B_\mu$ is parallel 
to a timelike Killing vector.
The modifications to $\Ph_g$ in this example
involve a rescaling of the Newton gravitational constant.

Examples with both $\Ph_B$ and $\Ph_q$ nonzero
can also be considered.
These incorporate direct charge couplings to matter,
but the static weak-field limit yields
modified gravitational and Coulomb potentials
given by Eq.\ \rf{Philambdasol}.
One simple example is the choice
$\Ph_B = - \Ph_q$.
Since $\Ph_g \ne \Ph_m$,
this $V_L$ model has a modified gravitational potential.
The solution for the bumblebee field 
has the form of a total derivative,
$\cE_\mu = \prt_\mu (t b \Ph_g)$.
In the limit $q \to 0$ and $\la \to 0$,
it provides an example of a solution
that with $\la = 0$ and hence $\Ph_B = 0$
has been identified in Ref.\ \cite{bb1}
as potentially flawed
due to the formation of shock discontinuities in $\cE_\mu$.
Whether or not $\la$ is zero,
this solution is unusual. 
Both the field strength $F_{\mu\nu}$ 
and the energy-momentum tensor $T_{\rm EM}^{\mu\nu}$ vanish
because the effective charge density $\rh_B$
associated with the Lagrange-multiplier field
cancels the matter charge density $\rh_q$.
Although the field strengths are zero,
the bumblebee field $\cE_\mu$ must be nonzero 
because the constraint \rf{lambdaconstraint}
implies $\cE_0 = b \Ph_g$,
which cannot vanish in the presence of gravity.
For a point charge the solution
$\cE_\mu$ does indeed contain a singularity,
but this is physically unremarkable 
as it merely reflects the usual $1/r$ dependence in $\Ph_g$.
The same behavior arises in the standard solutions
of Einstein-Maxwell theory in a gauge fixed
by Eq.\ \rf{lambdaconstraint},
to which the $V_L$ model is equivalent.

\subsection{Quadratic Lagrange-multiplier potential}
\label{Quadratic Lagrange-multiplier potential}

As a final example,
we consider in this subsection
the specific \ks\ bumblebee model \rf{ksbb}
with the quadratic Lagrange-multiplier potential \rf{VQ}
\beq
V = V_Q(\la, X) = \half \la (B_\mu g^{\mu\nu} B_{\nu} \pm b^2)^2,
\label{quadlagv}
\eeq
where $X$ is defined in Eq.\ \rf{XB}.
As before,
we adopt the mode expansion 
$B_\mu = b_\mu + \cE_\mu$ of Eq.\ \rf{epdown}
in a Minkowski background,
and where useful we assume weak fields $h_{\mu\nu}$ and $\cE_\mu$.

Bumblebee models with a quadratic Lagrange-multiplier potential
have not previously been considered in detail in the literature.
We introduce them here partly as a foil for the $V_L$ case,
in which the Lagrange multiplier plays a key role 
in the physics of the model. 
In contrast,
the Lagrange multiplier for the potential $V_Q$
decouples entirely from the classical dynamics.
The point is that 
variation with respect to $\la$ 
yields the constraint $X^2=0$,
which is equivalent to $X=0$ 
and forces the massive mode to vanish.
However,
the quadratic nature of $V_Q$
means that 
$\la$ always appears multiplied by $X$
in the gravitational and bumblebee field equations,
so the on-shell condition $X=0$ forces 
the field $\la$ to decouple.
Also,
the potential obeys both $V_Q =0$ and $V_Q^\prime = 0$ on shell,
so the bumblebee energy-momentum tensor and bumblebee current 
reduce to the standard expressions for electrodynamics 
in curved spacetime.
The equations of motion are therefore 
equivalent to those of Einstein-Maxwell electrodynamics
in the gauge $X=0$.
This correspondence holds both in the linearized limit
and in the full nonlinear theory.

Since the equations of motion generated 
by a theory with a quadratic Lagrange-multiplier potential
are the equations in the absence of the constraint
plus the constraint itself,
introducing this type of potential 
is equivalent at the classical level
to imposing the constraint by hand on the equations of motion.
The only distinction between the two approaches is the presence 
of the decoupled Lagrange-multiplier field.
Bumblebee models with the potential $V_Q$
therefore incorporate a variety of models 
in which constraints are imposed by hand.
For example,
Will and Nordtvedt have considered solutions 
involving a nonzero background value for a vector 
in bumblebee-type models with Lagrange density of the form \rf{bb}
but without a potential term 
\cite{wn}.
This approach produces equations of motion
identical to those of the corresponding $V_Q$ model. 

The $V_Q$ models are also related
to theories in which the constraint is substituted
into the action before varying,
although the correspondence is inexact. 
For example,
Nambu has investigated the Maxwell action in Minkowski spacetime
using the constraint $X=0$ for the purely timelike case
as a nonlinear gauge condition
substituted into the action prior to the variation
\cite{yn}.
This represents gauge fixing at the level of the action,
and it yields a total of four equations for the fields:
the original constraint and
three equations of motion from the variation.
However,
the corresponding $V_Q$ model 
yields five equations of motion instead,
one of which is the constraint.
The extra equation is the Gauss law,
which in the Nambu approach 
is imposed as a separate initial condition
that subsequently holds at all times 
by virtue of the three equations of motion and the constraint.

For the specific \ks\ bumblebee model \rf{ksbb}
with potential $V_Q$ in Eq.\ \rf{quadlagv},
the freely propagating modes can be found
by linearizing the gravitational and bumblebee equations of motion
\rf{geq} and \rf{Beq}
and the constraint $X=0$.
The linearization generates the same equations
as emerge for the $V_L$ model with potential \rf{linlagv}
in the limit $\la \to 0$.
The propagating modes 
consist of the usual graviton and photon in an axial gauge
\cite{bk},
as expected.
Similarly,
the weak static limit of the $V_Q$ model
with potential \rf{quadlagv}
produces equations identical at linear level
to those of the $V_L$ model
with potential \rf{linlagv}
in the limit $\la \to 0$.
The usual gravitational and electromagnetic 
Poisson equations therefore emerge,
and the correct Newton and Coulomb potentials hold
at the linearized level.

The exact correspondence between the static limit 
of the $V_Q$ bumblebee model and  
the Newton and Coulomb potentials 
involves the nonlinear constraint \rf{lmconst}.
The explicit forms of the solutions 
are therefore also nonlinear.
For example,
consider again the case of a point particle 
of mass $m$ and charge $q$ 
at rest at the origin
in the presence of a purely timelike vacuum value 
$b_\mu = (b,0,0,0)$.
The equation of motion for $\la$ is the constraint \rf{lmconst},
which represents a nonlinear condition
relating $\cE_\mu$ and $h_{\mu\nu}$.
Expanding this constraint through quadratic order gives 
\beq
\cE_0^2 +2 b \cE_0 - 2 b^2 \Ph_g - 4 b \Ph_m \cE_0 - \cE_j^2 
\approx 0 ,
\label{minquadV}
\eeq
where $\Ph_m$ obeys the conventional Poisson equation \rf{RN}
for the point source with mass and charge density
given by Eq.\ \rf{pointsource}.
In the transverse gauge,
the gravitational potential $\Ph_g$ obeys 
\beq
\vec \nabla^2 \Ph_g \approx 4\pi G (\rh_m + T_{\rm EM}^{00}) ,
\label{EMeq}
\eeq
where the energy-momentum tensor $T_{\rm EM}^{00}$
has the usual Einstein-Maxwell form \rf{TBBEM}.
The solution for $\Ph_g$ is therefore
the standard gravitational potential
for a static point charge 
in a curved spacetime in the weak-field limit.
At quadratic order,
the solution $\cE_\mu = (\cE_0, \cE_j)$ 
satisfying the constraint \rf{minquadV} 
and the field equations is found to be 
\bea
\cE_0 &\approx & b \Ph_g + \fr {3b} 2 \Ph_g^2 + \fr {t^2} {2b}
[\prt_j (\Ph_q + b \Ph_g)]^2 ,
\nonumber \\
\cE_j &\approx & t \prt_j (\Ph_q + b \Ph_g) 
+ 3bt \Ph_g \prt_j \Ph_g
\nonumber \\
&&
+ \fr {t^3} {3b} [\prt_k (\Ph_q + b \Ph_g)] \prt_j 
\prt_k (\Ph_q + b \Ph_g) ,
\label{quadep}
\eea
where $\Ph_q$ is the conventional Coulomb potential
obeying the Poisson equation \rf{fish}.
These expressions are both time dependent and nonlinear,
but they nonetheless generate 
the usual static electric field $\vec E$ 
and magnetic field $\vec B$ for a point charge.
Explicit calculation to quadratic order yields 
\beq
\vec E \approx - \vec \nabla \Ph_q ,
\qquad
\vec B \approx 0 ,
\label{EBfields2}
\eeq
which implies the usual form 
$\vec \nabla \cdot \vec E \approx \rh_q$
of the Gauss law.
From these equations,
we see again that Einstein-Maxwell theory is recovered
despite the absence of U(1) symmetry.
In effect, 
the nonlinear condition \rf{minquadV} 
plays the role of a nonlinear gauge-fixing condition
in a U(1)-invariant theory,
removing a degree of freedom and leaving only NG modes
that propagate as photons 
and generate the usual Coulomb potential in the weak-field limit.

\section{Summary}
\label{Summary}

In this paper,
we have investigated the properties of the massive modes 
that can emerge from
spontaneous local Lorentz and diffeomorphism breaking.
In Riemann spacetime,
no massive modes of the conventional Higgs type
can appear because covariant kinetic terms 
involve connections with derivatives. 
However,
an alternative form of the Higgs mechanism can occur instead,
in which massive modes originate from quadratic couplings 
in the potential $V$ inducing the symmetry breaking.

Section \ref{Massive Modes} 
provides an analysis of this alternative Higgs mechanism
in the general context of an arbitrary tensor field.
Both smooth potentials and Lagrange-multiplier potentials
are considered. 
Massive modes appear for a smooth potential
when excitations with $V^\prime \ne 0$ exist,
and they are formed as combinations of field and metric fluctuations.
For Lagrange-multiplier potentials
the massive modes are constrained to vanish,
but the Lagrange multiplier fields can play a related role. 
The propagation of the massive modes 
depends on the nature of the kinetic terms in the theory,
and the requirements of unitarity and ghost-free propagation
constrain possible models. 
Even if the massive modes are nonpropagating
they can influence gravitational phenomena through,
for example,
effects on the static gravitational potential.
These modifications are of potential interest
in alternative theories of gravity
and descriptions of phenomena 
such as dark matter or dark energy.

Following the general treatment,
we investigate in Sec.\ \ref{Bumblebee Models}
a broad class of theories 
called bumblebee models
that involve gravitationally coupled vector fields
with spontaneous local Lorentz and diffeomorphism breaking.
For arbitrary quadratic kinetic terms,
the Lagrange density is given in Eq.\ \rf{bb}.
Along with the symmetry-breaking potential $V$
and a matter sector,
this Lagrange density involves five parameters,
four of which can be linearly independent
in specific models.
A particularly attractive class of theories  
are the \ks\ bumblebee models,
which have kinetic term of the Maxwell form
and hence an additional constraint
that minimizes problems with unitarity and ghosts.
These models also offer candidate alternatives
to Einstein-Maxwell electrodynamics.

In a series of subsections and the associated appendix,
we provide some results valid for general bumblebee models.
The observer and particle forms of 
local Lorentz transformations and diffeomorphisms
are presented.
Using the vierbein formalism,
some decompositions of the bumblebee field and metric
are given that are suitable for 
the identification of Lorentz and diffeomorphism NG modes.
The effects of various choices of 
Lorentz and diffeomorphism gauges are described.
Alternative decompositions used in some of the literature
are also discussed,
in which the Lorentz NG modes are hidden
and only spacetime variables are used.

To provide explicit examples 
and to gain insight via a more detailed analysis,
we focus attention in Sec.\ \ref{Examples}
on the class of \ks\ bumblebee models.
The basic equations of motion and conservation laws are obtained,
and some properties of the bumblebee currents are considered.
The interpretation of these bumblebee models 
as theories of electromagnetism and gravity,
known as bumblebee electrodynamics,
is discussed. 
They contain four transverse massless modes,
two of which are massless gravitons
and two of which are massless photons,
along with a massive mode or Lagrange-multiplier mode.
When the massive mode or Lagrange multipler vanishes,
or in the limit of infinite mass,
conventional Einstein-Maxwell theory in an axial gauge
is recovered.

Section \ref{Examples} also contains subsections
considering in more detail various types of potentials,
including the smooth potential $V_S$
in Eq.\ \rf{smoothlagv},
the linear Lagrange-multiplier potential $V_L$
in Eq.\ \rf{linlagv},
and the quadratic Lagrange-multiplier potential $V_Q$
in Eq.\ \rf{quadlagv}.
For the $V_S$ model,
the gravitational and bumblebee equations of motion
are investigated to determine whether a physical massive mode
can propagate as a free field.
The sign of the squared mass term depends
on whether the bumblebee vacuum value $b_\mu$ 
is timelike or spacelike.
In the timelike case,
the massive mode is a ghost,
while in the spacelike case the squared mass has the usual sign.
However, 
in both cases the dispersion law is unconventional
and no localized physical solutions satisfying 
suitable asymptotic boundary conditions exist,
so the $V_S$ model has no freely propagating massive modes.
Nonetheless,
the massive mode has a physical impact as an auxiliary field,
acting as an additional source of energy-momentum density
and current density.
In the weak static limit,
for example,
the solutions to the equations of motion 
in the presence of mass and charge
describe modified Newton and Coulomb potentials
according to Eq.\ \rf{Phisol}.
These may be of phenomenological interest
in the context of dark matter 
and perhaps also dark energy.
The effects are controlled by the massive mode,
but its form is dynamically undetermined 
and must be imposed via initial conditions.

Many of the results obtained for the $V_S$ model 
apply also to the $V_L$ and $V_Q$ models 
that have Lagrange-multiplier fields.
For example,
all the models contain four massless propagating modes
behaving like gravitons and photons.
Although the $V_L$ and $V_Q$ models generate an additional constraint
that eliminates the massive mode on shell,
the Lagrange-multiplier field appears instead
as a extra degree of freedom
playing a similar role to that of the massive mode 
in the $V_S$ model.
The form of the Lagrange multiplier 
must be set by initial conditions,
and if it is chosen to vanish
then Einstein-Maxwell theory is recovered.
The key difference between the $V_L$ and $V_Q$ models
lies in the role of the Lagrange multiplier,
which can affect the physics as an auxiliary mode
in the $V_L$ model
but which decouples from the theory in the $V_Q$ model. 

In the context of bumblebee electrodynamics,
the massive mode or Lagrange-multiplier field
acts as an additional degree of freedom
relative to Einstein-Maxwell theory.
The extra freedom arises because the bumblebee model
has no U(1) gauge symmetry,
and the structure of the kinetic terms
implies that the freedom must be fixed as an initial condition.
In the absence of experimental evidence for a massive mode
or of guidance from an underlying theory,
the choice of initial condition is largely arbitrary.
A natural choice sets the massive mode or Lagrange multiplier field
to zero,
reducing the theory to Einstein-Maxwell electrodynamics
up to possible SME matter-sector couplings.
Bumblebee electrodynamics therefore provides 
a candidate alternative explanation
for the existence of massless photons,
based on the masslessness of NG modes
instead of the usual gauge symmetry. 
In any case,
the possibility of spontaneous breaking 
of local Lorentz and diffeomorphism symmetry
remains a promising avenue for exploring
physics emerging from the Planck scale.

\section*{Acknowledgments}

This work was supported in part
by DOE grant DE-FG02-91ER40661,
NASA grant NAG3-2914,
and NSF grant PHY-0554663.

\appendix

\section{Transformations}
\label{Transformations appendix}

In this appendix,
we provide explicit transformation formulae 
for the vierbein, bumblebee, and NG field components
under particle diffeomorphisms,
particle local Lorentz transformations,
observer general coordinate transformations,
and observer local Lorentz transformations.
The vierbein decomposition is given 
in Eq.\ \rf{vhch2},
the bumblebee decomposition is given
in Eq.\ \rf{Bvb2},
and the NG fields are defined in
Eq.\ \rf{promotion}.
The transformation formulae
can be deduced from these expressions
and from the behavior of the vacuum expectation values.
All the formulae given here are linearized 
assuming infinitesimal fields and transformation parameters.

\subsection{Particle transformations}
\label{Particle transformations}

Under infinitesimal particle diffeomorphisms,
\bea
b_\mu &\rightarrow& b_\mu ,
\nonumber\\ 
\et_{\mu\nu} &\rightarrow& \et_{\mu\nu} , 
\nonumber\\ 
\de^\mu_{\pt{\mu}\nu} &\rightarrow& \de^\mu_{\pt{\mu}\nu} , 
\nonumber\\ 
\be^{\rm t}_\mu &\rightarrow& \be^{\rm t}_\mu , 
\nonumber\\ 
\quad \be &\rightarrow& \be ,
\nonumber\\ 
h_{\mu\nu} &\rightarrow& h_{\mu\nu} - \prt_\mu \xi_\nu - 
\prt_\nu \xi_\mu ,
\nonumber\\ 
\ch_{\mu\nu} &\rightarrow& \ch_{\mu\nu} - \half (\prt_\mu \xi_\nu 
- \prt_\nu \xi_\mu ) 
\nonumber\\ 
\lvb \mu \nu \equiv \et_{\nu\al} \vb \mu \al &\rightarrow& \lvb \mu 
\nu - \prt_\mu \xi_\nu 
\nonumber\\ 
\uvb \mu \nu \equiv \et^{\nu\al} \ivb \mu \al &\rightarrow& \uvb \mu 
\nu + \prt^\nu \xi^\mu  ,
\nonumber\\ 
B_\mu &\rightarrow& B_\mu - (\prt_\mu \xi_\nu) b^\nu , 
\nonumber\\ 
B^\mu &\rightarrow& B^\mu + (\prt_\nu \xi^\mu) b^\nu ,
\nonumber\\ 
\Xi_\mu &\rightarrow& \Xi_\mu + \xi_\mu ,
\nonumber\\ 
\cE_{\mu\nu} &\rightarrow& \cE_{\mu\nu} .
\label{pdiffs}
\eea
Under infinitesimal local particle Lorentz transformations,
\bea
b_\mu &\rightarrow& b_\mu , 
\nonumber\\ 
\et_{\mu\nu} &\rightarrow& \et_{\mu\nu} , 
\nonumber\\ 
\de^\mu_{\pt{\mu}\nu} &\rightarrow& 
\de^\mu_{\pt{\mu}\nu} , 
\nonumber\\ 
\be^{\rm t}_\mu &\rightarrow& \be^{\rm t}_\mu + \ep_{\mu\nu} b^\nu, 
\nonumber\\ 
\be &\rightarrow& \be , 
\nonumber\\ 
h_{\mu\nu} &\rightarrow& h_{\mu\nu} ,
\nonumber\\ 
\ch_{\mu\nu} &\rightarrow& \ch_{\mu\nu} - \ep_{\mu\nu} 
\nonumber\\ 
\lvb \mu \nu \equiv \et_{\nu\al} \vb \mu \al &\rightarrow& \lvb \mu 
\nu - \ep_{\mu\nu} 
\nonumber\\ 
\uvb \mu \nu \equiv \et^{\nu\al} \ivb \mu \al &\rightarrow& \uvb \mu 
\nu - \ep^{\mu\nu} ,
\nonumber\\ 
B_\mu &\rightarrow& B_\mu ,
\nonumber\\ 
\Xi_\mu &\rightarrow& \Xi_\mu , 
\nonumber\\ 
\cE_{\mu\nu} &\rightarrow& \cE_{\mu\nu} + \ep_{\mu\nu} .
\label{pLTs}
\eea

\subsection{Observer transformations}
\label{Observer transformations}

The primary difference 
between observer and particle transformations
is that the components of vacuum-valued fields 
transform under observer transformations.
However,
additional differences arise when the distinction 
between spacetime and local indices is dropped.
In particular,
care is required in expressing observer transformation laws
for the metric and the vierbein.

Three versions of the Minkowski metric appear in the formalism:
the spacetime metric $\et_{\mu\nu}$,
the local metric $\et_{ab}$,
and the mixed metric $\et_{\mu a}$ used with the vierbein.
All three are numerically equal in cartesian coordinates,
but they behave differently under observer transformations. 
In the customary notation using only Greek indices, 
these three metrics must be labeled differently.
Since they represent vacuum values of the metric,
we write them as expectation values and define 
\bea
\vev{\et_{\mu\nu}}_\et &=& (\et_{ab})\vert_{ab \rightarrow \mu\nu} , 
\nonumber\\ 
\vev{\et_{\mu\nu}}_e &=& (\et_{\mu a})\vert_{a \rightarrow \nu} , 
\nonumber\\ 
\vev{\et_{\mu\nu}}_e &=& (\et_{\mu a})\vert_{a \rightarrow \nu} , 
\nonumber\\ 
\vev{\et_{\mu\nu}}_g &\equiv & \et_{\mu\nu} .
\label{mink3}
\eea
The inverse metrics are denoted with upper indices
and are defined so that
\bea
\vev{\et_{\mu\al}}_g \vev{\et^{\al\nu}}_g
&\approx & \vev{\et_{\mu\al}}_e \vev{\et^{\al\nu}}_e 
\approx 
\vev{\et_{\mu\al}}_\et \vev{\et^{\al\nu}}_\et
\approx  \de^\nu_{\pt{\mu}\mu} .
\nonumber\\ 
\label{invmink3}
\eea
In terms of these vacuum values,
the metric and its inverse involve infinitesimal excitations: 
\bea
g_{\mu\nu} &\approx & \vev{\et_{\mu\nu}}_g + h_{\mu\nu} , 
\nonumber\\ 
g^{\mu\nu} &\approx & \vev{\et^{\mu\nu}}_g - h^{\mu\nu} .
\label{g}
\eea

The transformation laws for the vierbein also depend
on the component basis. 
We define 
\bea
\vev{\lvb \mu \nu} &\approx & 
\vev{\et_{\mu\nu}}_e + \vev{\ch_{\mu\nu}} , 
\nonumber\\ 
\vev{\uvb \mu \nu} &\approx & 
\vev{\et^{\mu\nu}}_e + \vev{\ch^{\mu\nu}} , 
\nonumber\\ 
\vev{\ivb \mu \nu} &\approx & 
\vev{\et_{\nu\al}}_\et \vev{\uvb \mu \al} , 
\nonumber\\ 
\vev{\vb \mu \nu} &\approx & 
\vev{\et^{\nu\al}}_\et \vev{\lvb \mu \al} .
\label{obsvacs}
\eea
For example,
these vacuum values obey
\bea
\vev{\et_{\mu\nu}}_g &=& \vev{\lvb \mu \al} \vev{\lvb \nu \be} 
\vev{\et^{\al\be}}_\et 
\nonumber\\ 
&=& \vev{\et_{\mu\al}}_e \vev{\et_{\nu\be}}_e \vev{\et^{\al\be}}_\et ,
\label{gvbrel}
\eea
as expected from the relationship between the metric and the vierbein.
The vierbein field itself can be written 
\bea
\lvb \mu \nu \approx \vev{\et_{\mu\nu}}_e + \vev{\ch_{\mu\nu}}
+ \half h_{\mu\nu} + \ch_{\mu\nu} , 
\nonumber\\ 
\uvb \mu \nu \approx \vev{\et^{\mu\nu}}_e + \vev{\ch^{\mu\nu}}
- \half h^{\mu\nu} + \ch^{\mu\nu} .
\label{obsvb}
\eea

In much of the literature,
the vacuum value for $\vev{\ch_{\mu\nu}}$ is assumed to vanish
in cartesian coordinates.
However,
this choice is observer dependent.
Under either observer general coordinate transformations
or observer local Lorentz transformations,
a zero value of $\vev{\ch_{\mu\nu}}$ transforms 
into a nonzero value.
We therefore include the vacuum value
$\vev{\ch_{\mu\nu}}$ 
in the formulae below.

The transformation rules for infinitesimal observer
general coordinate transformations are
\bea
\vev{\et_{\mu\nu}}_g &\rightarrow& \vev{\et_{\mu\nu}}_g
- \prt_\mu \xi_\nu - \prt_\nu \xi_\mu , 
\nonumber\\ 
\vev{\et^{\mu\nu}}_g &\rightarrow& \vev{\et^{\mu\nu}}_g
+ \prt^\mu \xi^\nu + \prt^\nu \xi^\mu , 
\nonumber\\ 
\vev{\et_{\mu\nu}}_e &\rightarrow& \vev{\et_{\mu\nu}}_e
- \half (\prt_\mu \xi_\nu + \prt_\nu \xi_\mu) , 
\nonumber\\ 
\vev{\et^{\mu\nu}}_e &\rightarrow& \vev{\et^{\mu\nu}}_e
+ \half (\prt^\mu \xi^\nu + \prt^\nu \xi^\mu) , 
\nonumber\\ 
\vev{\et_{\mu\nu}}_\et &\rightarrow& \vev{\et_{\mu\nu}}_\et , 
\nonumber\\ 
\vev{\et^{\mu\nu}}_\et &\rightarrow& \vev{\et^{\mu\nu}}_\et , 
\nonumber\\ 
\vev{\ch_{\mu\nu}} &\rightarrow& \vev{\ch_{\mu\nu}}
- \half (\prt_\mu \xi_\nu - \prt_\nu \xi_\mu ), 
\nonumber\\ 
\vev{\ch^{\mu\nu}} &\rightarrow& \vev{\ch^{\mu\nu}}
- \half (\prt^\mu \xi^\nu - \prt^\nu \xi^\mu ), 
\nonumber\\ 
\vev{\lvb \mu \nu} &\rightarrow& \vev{\lvb \mu \nu} - \prt_\mu 
\xi_\nu , 
\nonumber\\ 
\vev{\uvb \mu \nu} &\rightarrow& \vev{\uvb \mu \nu} + \prt^\nu 
\xi^\mu , 
\nonumber\\ 
b_\mu &\rightarrow& b_\mu - (\prt_\mu \xi_\nu) b^\nu, 
\nonumber\\ 
b^\mu = \vev{\et^{\mu\nu}}_g b_\nu &\rightarrow& b^\mu + 
(\prt_\nu \xi^\mu) b^\nu , 
\nonumber\\ 
\be^{\rm t}_\mu &\rightarrow& \be^{\rm t}_\mu , 
\nonumber\\ 
\be &\rightarrow& \be , 
\nonumber\\ 
h_{\mu\nu} &\rightarrow& h_{\mu\nu} , 
\nonumber\\ 
\ch_{\mu\nu} &\rightarrow& \ch_{\mu\nu} ,
\nonumber\\ 
\lvb \mu \nu &\rightarrow& \lvb \mu \nu - \prt_\mu \xi_\nu , 
\nonumber\\ 
\uvb \mu \nu &\rightarrow& \uvb \mu \nu + \prt^\nu \xi^\mu ,
\nonumber\\ 
B_\mu &\rightarrow& B_\mu - (\prt_\mu \xi_\nu ) b^\nu , 
\nonumber\\ 
B^\mu &\rightarrow& B^\mu + (\prt_\nu \xi^\mu ) b^\nu ,
\nonumber\\ 
\Xi_\mu &\rightarrow& \Xi_\mu ,
\nonumber\\ 
\cE_{\mu\nu} &\rightarrow& \cE_{\mu\nu} .
\label{odiffs}
\eea

The formulae for infinitesimal observer 
local Lorentz transformations are
\bea
\vev{\et_{\mu\nu}}_g &\rightarrow& \vev{\et_{\mu\nu}}_g , 
\nonumber\\ 
\vev{\et_{\mu\nu}}_e &\rightarrow& \vev{\et_{\mu\nu}}_e , 
\nonumber\\ 
\vev{\et_{\mu\nu}}_\et &\rightarrow& \vev{\et_{\mu\nu}}_\et  , 
\nonumber\\ 
\vev{\ch_{\mu\nu}} &\rightarrow& 
\vev{\ch_{\mu\nu}} - \ep_{\mu\nu} , 
\nonumber\\ 
\vev{\lvb \mu \nu} &\rightarrow& 
\vev{\lvb \mu \nu}  - \ep_{\mu\nu} , 
\nonumber\\ 
b_\mu &\rightarrow& b_\mu + \ep_{\mu\nu} b^\nu, 
\nonumber\\ 
\be^{\rm t}_\mu &\rightarrow& \be^{\rm t}_\mu , 
\nonumber\\ 
\be &\rightarrow& \be , 
\nonumber\\ 
h_{\mu\nu} &\rightarrow& h_{\mu\nu} , 
\nonumber\\ 
\ch_{\mu\nu} &\rightarrow& \ch_{\mu\nu} ,
\nonumber\\ 
\lvb \mu \nu &\rightarrow& \lvb \mu \nu - \ep_{\mu\nu} ,
\nonumber\\ 
B_\mu &\rightarrow& B_\mu ,
\nonumber\\ 
\Xi_\mu &\rightarrow& \Xi_\mu , 
\nonumber\\ 
\cE_{\mu\nu} &\rightarrow& \cE_{\mu\nu} .
\label{oLTs}
\eea
In these expressions,
indices can be raised and lowered using
$\vev{\et_{\mu\nu}}_\et$.

For some purposes,
it may be desirable 
for an observer transformation
to take the appearance 
of the inverse of a particle transformation.
This can be achieved by changing
the signs of the parameters,
$\xi_\mu \rightarrow - \xi_\mu$ 
and $\ep_{\mu\nu} \rightarrow - \ep_{\mu\nu}$,
either in the formulae for the observer transformations
or in those for the particle transformations.
For example,
with signs chosen appropriately
the bumblebee field $B_\mu$ can be arranged to be invariant,
$B_\mu \rightarrow B_\mu$,
under the composite transformation
consisting of a particle diffeomorphism followed by  
the corresponding inverse general coordinate transformation.
However,
the reader is cautioned that invariance of this type may fail 
for the vacuum value and excitations of $B_\mu$.
For instance,
the vacuum value $b_\mu$ 
is invariant under particle diffeomorphisms
but transforms as ordinary components of a vector 
under observer general coordinate transformations.


\begin{thebibliography}{99}

\bibitem{ng}
Y.\ Nambu,
Phys.\ Rev.\ Lett.\ {\bf 4}, 380 (1960);
J.\ Goldstone,
Nuov.\ Cim.\ {\bf 19}, 154 (1961);
J.\ Goldstone, A.\ Salam, and S.\ Weinberg,
Phys.\ Rev.\ {\bf 127}, 965 (1962).

\bibitem{hm}
P.W.\ Anderson,
Phys.\ Rev.\ {\bf 130}, 439 (1963);
P.W.\ Higgs,
Phys.\ Lett.\ {\bf 12}, 132 (1964);
F.\ Englert and R.\ Brout,
Phys.\ Rev.\ Lett.\ {\bf 13}, 321 (1964);
G.S.\ Guralnik, C.R.\ Hagen, and T.W.B.\ Kibble,
Phys.\ Rev.\ Lett.\ {\bf 13}, 585 (1964).

\bibitem{ks}
V.A.\ Kosteleck\'y and S.\ Samuel,
Phys.\ Rev.\ D {\bf 40}, 1886 (1989);
Phys.\ Rev.\ Lett.\ {\bf 63}, 224 (1989).

\bibitem{akgrav}
V.A.\ Kosteleck\'y,
Phys.\ Rev.\ D {\bf 69}, 105009 (2004).

\bibitem{bk}
R.\ Bluhm and V.A.\ Kosteleck\'y,
Phys.\ Rev.\ D {\bf 71}, 065008 (2005).

\bibitem{ksp}
V.A.\ Kosteleck\'y and S.\ Samuel,
Phys.\ Rev.\ D {\bf 39}, 683 (1989);
V.A.\ Kosteleck\'y and R.\ Potting,
Nucl.\ Phys.\ B {\bf 359}, 545 (1991).

\bibitem{ncqed}
See, for example,
I.\ Mocioiu, M.\ Pospelov, and R.\ Roiban,
Phys.\ Lett.\ B {\bf 489}, 390 (2000);
S.M.\ Carroll \etal,
Phys.\ Rev.\ Lett.\ {\bf 87}, 141601 (2001);
Z.\ Guralnik, R.\ Jackiw, S.Y.\ Pi, and A.P.\ Polychronakos,
Phys.\ Lett.\ B {\bf 517}, 450 (2001);
C.E.\ Carlson, C.D.\ Carone, and R.F.\ Lebed,
Phys.\ Lett.\ B {\bf 518}, 201 (2001);
A.\ Anisimov, T.\ Banks, M.\ Dine, and M.\ Graesser,
Phys.\ Rev.\ D {\bf 65}, 085032 (2002);
A.\ Das, J.\ Gamboa, J.\ Lopez-Sarrion, and F.A.\ Schaposnik,
Phys.\ Rev.\ D {\bf 72}, 107702 (2005).

\bibitem{spacetimevarying}
O.\ Bertolami, R.\ Lehnert, R.\ Potting, and A.\ Ribeiro,
Phys.\ Rev.\ D {\bf 69}, 083513 (2004);
V.A.\ Kosteleck\'y, R.\ Lehnert, and M.\ Perry,
Phys.\ Rev.\ D {\bf 68}, 123511 (2003);
R.\ Jackiw and S.-Y.\ Pi,
Phys.\ Rev.\ D {\bf 68}, 104012 (2003).

\bibitem{qg}
See, for example,
G.\ Amelino-Camelia, C.\ L\"ammerzahl, A.\ Macias, and H.\ M\"uller,
AIP Conf.\ Proc.\ {\bf 758}, 30 (2005);
N.E.\ Mavromatos,
Lect.\ Notes Phys.\ {\bf 669}, 245 (2005);
H.A.\ Morales-Tecotl and L.F.\ Urrutia,
AIP Conf.\ Proc.\ {\bf 857}, 205 (2006);
Y.\ Bonder and D.\ Sudarsky,
arXiv:0709.0551.

\bibitem{fn}
C.D.\ Froggatt and H.B.\ Nielsen,
hep-ph/0211106.

\bibitem{bj}
J.D.\ Bjorken,
Phys.\ Rev.\ D {\bf 67}, 043508 (2003).

\bibitem{brane}
See, for example,
C.P.\ Burgess, J.\ Cline, E.\ Filotas,
J.\ Matias, and G.D.\ Moore,
JHEP {\bf 0203}, 043 (2002);
A.R.\ Frey,
JHEP {\bf 0304}, 012 (2003);
J.\ Cline and L.\ Valc\'arcel,
JHEP {\bf 0403}, 032 (2004).

\bibitem{susy}
M.\ Berger and V.A.\ Kosteleck\'y,
Phys.\ Rev.\ D {\bf 65}, 091701(R) (2002);
P.A.\ Bolokhov, S.G.\ Nibbelink, and M.\ Pospelov,
Phys.\ Rev.\ D {\bf 72}, 015013 (2005).

\bibitem{modgrav1}
V.A.\ Kosteleck\'y and S.\ Samuel,
Phys.\ Rev.\ D {\bf 42}, 1289 (1990);
Phys.\ Rev.\ Lett.\ {\bf 66}, 1811 (1991).

\bibitem{modgrav2}
N.\ Arkani-Hamed, H.-C.\ Cheng, M.\ Luty, and S.\ Mukohyama,
JHEP {\bf 0405}, 074 (2004);
N.\ Arkani-Hamed, H.-C.\ Cheng, M.\ Luty, and J.\ Thaler,
JHEP {\bf 0507}, 029 (2005);
D.S.\ Gorbunov and S.M.\ Sibiryakov,
JHEP {\bf 0509}, 082 (2005);
H.-C.\ Cheng, M.\ Luty, S.\ Mukohyama, and J.\ Thaler,
JHEP {\bf 0605}, 076 (2006);
G.\ Dvali, O.\ Pujolas, and M.\ Redi,
Phys.\ Rev.\ D {\bf 76}, 044028 (2007).

\bibitem{modgrav3}
M.V.\ Libanov and V.A.\ Rubakov,
JHEP {\bf 0508}, 001 (2005);
S.\ Dubovsky, P.\ Tinyakov, and M.\ Zaldarriaga,
arXiv:0706.0288.

\bibitem{kp}
V.A.\ Kosteleck\'y and R.\ Potting,
Phys.\ Rev.\ D {\bf 51}, 3923 (1995).

\bibitem{ck}
D.\ Colladay and V.A.\ Kosteleck\'y,
Phys.\ Rev.\ D {\bf 55}, 6760 (1997);
Phys.\ Rev.\ D {\bf 58}, 116002 (1998).

\bibitem{cpt07}
For recent reviews of various experimental
and theoretical approaches to Lorentz violation
see, for example,
V.A.\ Kosteleck\'y, ed.,
{\it CPT and Lorentz Symmetry I-IV},
World Scientific, Singapore, 1999-2008;
R.\ Bluhm,
Lect.\ Notes Phys.\ {\bf 702}, 191 (2006)
[hep-ph/0506054];
D.M.\ Mattingly,
Living Rev.\ Rel.\ {\bf 8}, 5 (2005).

\bibitem{kr}
A tabulation of results is given in 
V.A.\ Kosteleck\'y and N.\ Russell,
arXiv:0801.0287.

\bibitem{photonexpt}
J.\ Lipa \etal,
Phys.\ Rev.\ Lett.\ {\bf 90}, 060403 (2003);
H.\ M\"uller \etal,
Phys.\ Rev.\ Lett.\ {\bf 91}, 020401 (2003);
P.\ Wolf \etal,
Gen.\ Rel.\ Grav.\ {\bf 36}, 2351 (2004);
Phys.\ Rev.\ D {\bf 70}, 051902 (2004);
M.\ Tobar \etal,
Phys.\ Rev.\ D {\bf 71}, 025004 (2005);
S.\ Herrmann \etal,
Phys.\ Rev.\ Lett.\ {\bf 95}, 150401 (2005);
P.L.\ Stanwix et al., 
Phys.\ Rev.\ D {\bf 74}, 081101 (R) (2006);
M.\ Hohensee et al., 
Phys.\ Rev.\ D {\bf 71}, 025004 (2007);
H.\ M\"uller \etal,
Phys.\ Rev.\ Lett.\ {\bf 99}, 050401 (2007);
S.\ Reinhardt et al., 
Nature Physics {\bf 3}, 861 (2007);
S.M.\ Carroll, G.B.\ Field, and R.\ Jackiw,
Phys.\ Rev.\ D {\bf 41}, 1231 (1990);
R.\ Jackiw and V.A.\ Kosteleck\'y,
Phys.\ Rev.\ Lett.\ {\bf 82}, 3572 (1999);
V.A.\ Kosteleck\'y and M.\ Mewes,
Phys.\ Rev.\ Lett.\ {\bf 87}, 251304 (2001);
Phys.\ Rev.\ D {\bf 66}, 056005 (2002);
Phys.\ Rev.\ Lett.\ {\bf 97}, 140401 (2006);
Phys.\ Rev.\ Lett.\ {\bf 99}, 011601 (2007);
Q.G.\ Bailey and V.A.\ Kosteleck\'y,
Phys.\ Rev.\ D {\bf 70}, 076006 (2004);
C.D.\ Carone, M.\ Sher, M.\ Vanderhaeghen
Phys.\ Rev.\ D {\bf 74}, 077901 (2006);
B.\ Altschul,
Phys.\ Rev.\ Lett.\ {\bf 96}, 201101 (2006);
Phys.\ Rev.\ Lett.\ {\bf 98}, 041603 (2007);
Phys.\ Rev.\ D {\bf 75}, 105003 (2007).

\bibitem{eexpt}
H.\ Dehmelt \etal,
Phys.\ Rev.\ Lett.\ {\bf 83}, 4694 (1999);
R.\ Mittleman \etal,
Phys.\ Rev.\ Lett.\ {\bf 83}, 2116 (1999);
G.\ Gabrielse \etal,
Phys.\ Rev.\ Lett.\ {\bf 82}, 3198 (1999);
R.\ Bluhm \etal,
Phys.\ Rev.\ Lett.\ {\bf 82}, 2254 (1999);
Phys.\ Rev.\ Lett.\ {\bf 79}, 1432 (1997);
Phys.\ Rev.\ D {\bf 57}, 3932 (1998);
D.\ Colladay and V.A.\ Kosteleck\'y,
Phys.\ Lett.\ B {\bf 511}, 209 (2001);
B.\ Altschul,
Phys.\ Rev.\ D {\bf 74}, 083003 (2006);
G.M.\ Shore,
Nucl.\ Phys.\ B {\bf 717}, 86 (2005).

\bibitem{eexpt2}
B.\ Heckel \etal,
Phys.\ Rev.\ Lett.\ {\bf 97}, 021603 (2006);
L.-S.\ Hou, W.-T.\ Ni, and Y.-C.M.\ Li,
Phys.\ Rev.\ Lett.\ {\bf 90}, 201101 (2003);
R.\ Bluhm and V.A.\ Kosteleck\'y,
Phys.\ Rev.\ Lett.\ {\bf 84}, 1381 (2000).

\bibitem{eexpt3}
H.\ M\"uller, S.\ Herrmann, A.\ Saenz, A.\ Peters,
and C.\ L\"ammerzahl,
Phys. Rev. D {\bf 70}, 076004 (2004);
H.\ M\"uller,
Phys.\ Rev.\ D {\bf 71}, 045004 (2005).

\bibitem{ccexpt}
D.\ Bear \etal,
Phys.\ Rev.\ Lett.\ {\bf 85}, 5038 (2000);
D.F.\ Phillips \etal,
Phys.\ Rev.\ D {\bf 63}, 111101 (2001);
M.A.\ Humphrey \etal,
Phys.\ Rev.\ A {\bf 68}, 063807 (2003);
F.\ Can\`e \etal,
Phys.\ Rev.\ Lett.\ {\bf 93}, 230801 (2004);
P.\ Wolf \etal,
Phys.\ Rev.\ Lett.\ {\bf 96}, 060801 (2006);
M.\ Romalis,
in
{\it CPT and Lorentz Symmetry IV},
Ref.\ \cite{cpt07};
V.A.\ Kosteleck\'y and C.D.\ Lane,
Phys.\ Rev.\ D {\bf 60}, 116010 (1999);
J.\ Math.\ Phys.\ {\bf 40}, 6245 (1999);
C.D.\ Lane,
Phys.\ Rev.\ D {\bf 72}, 016005 (2005);
D.\ Colladay and P.\ McDonald,
Phys.\ Rev.\ D {\bf 73}, 105006 (2006).

\bibitem{spaceexpt}
R.\ Bluhm \etal,
Phys.\ Rev.\ Lett.\ {\bf 88}, 090801 (2002);
Phys.\ Rev.\ D {\bf 68}, 125008 (2003).

\bibitem{bnsyn}
O.\ Bertolami \etal,
Phys.\ Lett.\ B {\bf 395}, 178 (1997);
G.\ Lambiase,
Phys.\ Rev.\ D {\bf 72}, 087702 (2005);
J.M.\ Carmona, J.L.\ Cort\'es, A.\ Das,
J.\ Gamboa, and F.\ M\'endez,
Mod.\ Phys.\ Lett.\ {\bf A21}, 883 (2006).

\bibitem{hadronexpt}
KTeV Collaboration,
H.\ Nguyen,
in
{\it CPT and Lorentz Symmetry II}, Ref.\ \cite{cpt07}
[hep-ex/0112046];
KLOE Collaboration,
A.\ Di Domenico,
in
{\it CPT and Lorentz Symmetry IV}, Ref.\ \cite{cpt07};
OPAL Collaboration,
R.\ Ackerstaff \etal,
Z.\ Phys.\ C {\bf 76}, 401 (1997);
DELPHI Collaboration,
M.\ Feindt \etal,
preprint DELPHI 97-98 CONF 80 (1997);
BELLE Collaboration,
K.\ Abe \etal,
Phys.\ Rev.\ Lett.\ {\bf 86}, 3228 (2001);
BaBar Collaboration,
B.\ Aubert
\etal,
Phys.\ Rev.\ Lett.\ {\bf 92}, 142002 (2004);
hep-ex/0607103;
arXiv:0711.2713;
FOCUS Collaboration,
J.M.\ Link \etal,
Phys.\ Lett.\ B {\bf 556}, 7 (2003);
V.A.\ Kosteleck\'y,
Phys.\ Rev.\ Lett.\ {\bf 80}, 1818 (1998);
Phys.\ Rev.\ D {\bf 61}, 016002 (2000);
Phys.\ Rev.\ D {\bf 64}, 076001 (2001);
N.\ Isgur \etal,
Phys.\ Lett.\ B {\bf 515}, 333 (2001).

\bibitem{muexpt}
$g$--2 Collaboration,
G.W.\ Bennett \etal,
Phys.\ Rev.\ Lett., in press [arXiv:0709.4670];
V.W.\ Hughes \etal,
Phys.\ Rev.\ Lett.\ {\bf 87}, 111804 (2001);
R.\ Bluhm \etal,
Phys.\ Rev.\ Lett.\ {\bf 84}, 1098 (2000).

\bibitem{nuexpt}
LSND Collaboration,
L.B.\ Auerbach \etal,
Phys.\ Rev.\ D {\bf 72}, 076004 (2005);
M.D.\ Messier (SK),
in {\it CPT and Lorentz Symmetry III},
Ref.\ \cite{cpt07};
B.J.\ Rebel and S.F.\ Mufson (MINOS),
in {\it CPT and Lorentz Symmetry IV},
Ref.\ \cite{cpt07};
V.A.\ Kosteleck\'y and M.\ Mewes,
Phys.\ Rev.\ D {\bf 69}, 016005 (2004);
Phys.\ Rev.\ D {\bf 70}, 031902(R) (2004);
Phys.\ Rev.\ D {\bf 70}, 076002 (2004);
T.\ Katori, V.A.\ Kosteleck\'y, and R.\ Tayloe,
Phys.\ Rev.\ D {\bf 74}, 105009 (2006);
V.\ Barger, D.\ Marfatia, and K.\ Whisnant,
Phys.\ Lett.\ B {\bf 653}, 267 (2007).

\bibitem{higgs}
D.L.\ Anderson, M.\ Sher, and I.\ Turan,
Phys.\ Rev.\ D {\bf 70}, 016001 (2004);
E.O.\ Iltan,
Mod.\ Phys.\ Lett.\ A {\bf 19}, 327 (2004).

\bibitem{gravexpt}
J.B.R.\ Battat, J.F.\ Chandler, and C.W.\ Stubbs,
Phys.\ Rev.\ Lett.\ {\bf 99}, 241103 (2007);
H.\ M\"uller \etal,
Phys.\ Rev.\ Lett.\ {\bf 100}, 031101 (2008);
W.M.\ Jensen, S.M.\ Lewis, and J.C.\ Long,
in {\it CPT and Lorentz Symmetry IV},
Ref.\ \cite{cpt07}.

\bibitem{bak06}
Q.G.\ Bailey and V.A.\ Kosteleck\'y,
Phys.\ Rev.\ D {\bf 74}, 045001 (2006).

\bibitem{uk}
R.\ Utiyama,
Phys.\ Rev.\ {\bf 101}, 1597 (1956);
T.W.B.\ Kibble,
J.\ Math.\ Phys.\ {\bf 2}, 212 (1961).

\bibitem{hmnonlv}
See, for example,
R.\ Percacci, 
Nucl.\ Phys.\ B {\bf 353}, 271 (1991).

\bibitem{kpgr}
V.A.\ Kosteleck\'y and R.\ Potting,
Gen.\ Rel.\ Grav.\ {\bf 37}, 1675 (2005).

\bibitem{fp}
M.\ Fierz, 
Helv.\ Phys.\ Acta {\bf 12}, 3 (1939); 
M.\ Fierz and W.\ Pauli, 
Proc.\ Roy.\ Soc.\ {\bf 173}, 211 (1939).

\bibitem{vdvz}
H.\ van Dam and M.\ Veltman,
Nucl.\ Phys.\ B {\bf 22}, 397 (1970);
V.I.\ Zakharov,
JEPT Lett.\ {\bf 12}, 312 (1970).

\bibitem{gyb}
G.Yu.\ Bogoslovsky,
SIGMA {\bf 1}, 017 (2005);
Phys.\ Lett.\ A {\bf 350}, 5 (2006);
X.\ Li and Z.\ Chang,
arXiv:0711.1934v1.

\bibitem{ems}
See, for example,
J.W.\ Elliott, G.D.\ Moore, and H.\ Stoica,
JHEP {\bf 0508}, 066 (2005);
M.D.\ Seifert
Phys.\ Rev.\ D {\bf 76}, 064002 (2007).

\bibitem{torsion}
See, for example,
F.W.\ Hehl, P.\ von der Heyde, G.D.\ Kerlick, and J.M.\ Nester,
Rev.\ Mod.\ Phys.\ {\bf 48}, 393 (1976);
I.L.\ Shapiro,
Phys.\ Rep.\ {\bf 357}, 113 (2002);
V.A.\ Kosteleck\'y, N.\ Russell, and J.D.\ Tasson,
Phys.\ Rev.\ Lett., in press [arXiv:0712.4393].

\bibitem{wn}
C.M.\ Will and K.\ Nordtvedt,
Astrophys.\ J.\ {\bf 177}, 757 (1972);
R.W.\ Hellings and K.\ Nordtvedt,
Phys.\ Rev.\ D {\bf 7}, 3593 (1973).
See also 
C.M.\ Will,
\it Theory and Experiment in Gravitational Physics, \rm
Cambridge University Press, Cambridge, 1993.

\bibitem{kleh}
V.A.\ Kosteleck\'y and R.\ Lehnert,
Phys.\ Rev.\ D {\bf 63}, 065008 (2001).

\bibitem{baak}
B.\ Altschul and V.A.\ Kosteleck\'y,
Phys.\ Lett.\ B {\bf 628}, 106 (2005).

\bibitem{bb1}
C.\ Eling, T.\ Jacobson, and D.\ Mattingly, gr-qc/0410001.

\bibitem{bb2}
C.\ Eling and T.\ Jacobson,
Class.\ Quant.\ Grav.\ {\bf 23}, 5643 (2006);
C.\ Eling,
Phys.\ Rev.\ D {\bf 73}, 084026 (2006);
B.Z.\ Foster,
Phys.\ Rev.\ D {\bf 73}, 104012 (2006).

\bibitem{cli}
S.M.\ Carroll and E.A.\ Lim,
Phys.\ Rev.\ D {\bf 70}, 123525 (2004).

\bibitem{bmg}
B.M.\ Gripaios,
JHEP {\bf 0410}, 069 (2004).

\bibitem{gjw}
M.L.\ Graesser, A.\ Jenkins, and M.B.\ Wise,
Phys.\ Lett.\ B {\bf 613}, 5 (2005).

\bibitem{cfn}
J.L.\ Chkareuli, C.D.\ Froggatt, and H.B.\ Nielsen, 
hep-th/0610186.

\bibitem{bb3}
J.D.\ Bjorken,
hep-th/0111196;
P.\ Kraus and E.T.\ Tomboulis,
Phys.\ Rev.\ D {\bf 66}, 045015 (2002);
J.W.\ Moffat,
Intl.\ J.\ Mod.\ Phys.\ D {\bf 12}, 1279 (2003);
E.A.\ Lim,
Phys.\ Rev.\ D {\bf 71}, 063504 (2005);
C.\ Heinicke, P.\ Baekler, and F.W.\ Hehl,
Phys.\ Rev.\ D {\bf 72}, 025012 (2005);
O.\ Bertolami and J.\ Paramos,
Phys.\ Rev.\ D {\bf 72}, 044001 (2005);
S.\ Kanno and J.\ Soda,
Phys.\ Rev.\ D {\bf 74}, 063505 (2006);
L.\ Ackerman, S.M.\ Carroll, and M.B.\ Wise,
Phys.\ Rev.\ D {\bf 75}, 083502 (2007);
J.L.\ Chkareuli, C.D.\ Froggatt, J.G.\ Jejelava, and H.B.\ Nielsen,
arXiv:0710.3479;
A.\ Halle and H.\ Zhao,
arXiv:0711.0958.

\bibitem{dhfb}
P.A.M.\ Dirac,
Proc.\ R.\ Soc.\ Lon.\ {\bf A209}, 291, (1951);
W.\ Heisenberg,
Rev.\ Mod.\ Phys.\ {\bf 29}, 269 (1957);
P.G.O.\ Freund,
Acta Phys.\ Austriaca {\bf 14}, 445 (1961);
J.D.\ Bjorken,
Ann.\ Phys.\ {\bf 24}, 174 (1963).

\bibitem{yn}
Y.\ Nambu,
Prog.\ Theor.\ Phys.\ Suppl.\ Extra 190 (1968).

\bibitem{cj}
M.A.\ Clayton, gr-qc/0104103.

\bibitem{renorm}
V.A.\ Kosteleck\'y, C.D.\ Lane, and A.G.M.\ Pickering,
Phys.\ Rev.\ D {\bf 65}, 056006 (2002);
V.A.\ Kosteleck\'y and A.G.M.\ Pickering,
Phys.\ Rev.\ Lett.\ {\bf 91}, 031801 (2003);
G.\ de Berredo-Peixoto and I.L.\ Shapiro,
Phys.\ Lett.\ B {\bf 642}, 153 (2006);
D.\ Colladay and P.\ McDonald,
Phys.\ Rev.\ D {\bf 75}, 105002 (2007);
arXiv:0712.2055.

\end{thebibliography}
\end{document}